\documentclass[reprint,onecolumn,aps,showpacs,nofootinbib,superscriptaddress,notitlepage]{revtex4-2}
\pdfoutput=1
\usepackage[normalem]{ulem}
\usepackage{graphicx}
\usepackage{psfrag}
\usepackage{caption}
\usepackage{subcaption}
\usepackage{float}
\usepackage{amsmath}
\usepackage{amssymb}
\usepackage{amsthm}
\usepackage{mathtools}
\usepackage{dsfont}
\usepackage{xcolor}
\usepackage{comment}
\usepackage{caption}
\usepackage{geometry}
\usepackage{lineno}
\usepackage{ulem}

\newcommand{\be}{\begin{equation}}
\newcommand{\ee}{\end{equation}}

\def\bc{\begin{center}}
\def\ec{\end{center}}
\def\bea{\begin{eqnarray}}
\def\eea{\end{eqnarray}}

\newcommand{\deletetext}[1]{\iffalse{{\color{red}{#1}}}\fi}
\newcommand{\newtext}[1]{{\color{black}{#1}}}

\begin{document} %All text i dokumentet hamnar mellan dessa taggar, allt ovanför är formatering av dokumentet

\title{Graph-combinatorial approach for large deviations of Markov chains}

\author{Giorgio Carugno}
\email{giorgio.carugno@kcl.ac.uk}
\affiliation{Department of Mathematics, King’s College London, Strand, London WC2R 2LS, UK}

\author{Pierpaolo Vivo}
\email{pierpaolo.vivo@kcl.ac.uk }
\affiliation{Department of Mathematics, King’s College London, Strand, London WC2R 2LS, UK}

\author{Francesco Coghi}
\email{francesco.coghi@su.se}
\affiliation{Nordita, KTH Royal Institute of Technology and Stockholm University, Hannes Alfvéns väg 12, SE-106 91 Stockholm, Sweden}

\date{\today}

%\linenumbers

\begin{abstract}
We consider discrete-time Markov chains and study large deviations of the pair empirical occupation measure, which is useful to compute fluctuations of pure-additive and jump-type observables. We provide an exact expression for the finite-time moment generating function, \newtext{which is split in cycles and paths contributions}, and scaled cumulant generating function of the pair empirical occupation measure via a graph-combinatorial approach. The expression obtained allows us to give a physical interpretation of interaction and entropic terms, and of the Lagrange multipliers, and may serve as a starting point for sub-leading asymptotics. \newtext{We illustrate the use of the method for a simple two-state Markov chain.}
\end{abstract}

\maketitle

\section{Fluctuations for discrete-time Markov chains in the large deviations regime}

Markov chains are widely-used stochastic models of in and out-of-equilibrium physical systems. We consider a discrete-time ergodic Markov chain $X = \left( X_{\ell} \right)_{\ell=1}^{n+1} = (X_1,X_2,\dots,X_{n+1})$ evolving in a finite discrete state space $\Gamma$ of $N$ states according to the (irreducible and aperiodic) transition matrix $\Pi$. The matrix $\Pi$ characterises the probability of going from a state $X_{\ell}=i$ at time $\ell$ to a state $X_{\ell+1}=j$ at time $\ell+1$. We will use the index $\ell$ to refer to time and the indices $i$ and $j$ to refer to general states of the state space.

In this setting, one and two-point observables having the general form
\begin{equation}
\label{eq:Cost}
C_n = \frac{1}{n} \sum_{\ell=1}^n f(X_{\ell},X_{\ell+1}) \, ,
\end{equation}
where $f$ is any function that may depend both on the starting and landing state, are of fundamental importance to characterise the typical and fluctuating behaviour of the associated physical systems. [Notice that by taking $f(i,j)=g(i)$, $C_n$ in \eqref{eq:Cost} can also cover the case of purely time-additive observables\footnote{This marginalisation only works for discrete-time processes.}.] Just to give an example, the observable in \eqref{eq:Cost} can represent the number of transitions over $[1,n]$ in a particular subset of the state space \cite{Chetrite2015}, obtained by fixing $f=\mathbf{1}_{\Delta}$, with $\Delta$ the characteristic function of the subset. Furthermore, in certain contexts, $C_n$ can also express heat \cite{Sekimoto2010}, two-point correlation functions, activities \cite{Maes2008,Baiesi2009,Gutierrez2021,Dechant2021}, particle and energy currents \cite{Derrida2007}, efficiency \cite{Verley2014,Verley2014a,Gingrich2014,Manikandan2019,Coghi2020}, entropy production \cite{Lebowitz1999,Mehl2008,Dechant2021}, and many others.

To study fluctuations of $C_n$, the probabilistic theory of large deviations and, in particular, the Donsker--Varadhan approach may be used as it offers analytical and numerical methods to calculate the large deviation (or rate) function
\begin{equation}
    \label{eq:Rate}
    I(c) = - \lim_{n \rightarrow \infty} \frac{1}{n} \ln \mathbb{P}(C_n=c)
\end{equation}
characterising the time-leading exponential behaviour -- provided there is one -- of the probability distribution $\mathbb{P}(C_n=c)$ \cite{Deuschel1989,DenHollander2000,Touchette2009,Dembo2010,Touchette2013,Chetrite2015,Chetrite2015a}. The rate function in \eqref{eq:Rate} is always positive and measures the extent of the fluctuations of $C_n$ around its typical value $c^*$, which, for ergodic Markov chains, is the unique zero of $I$ \cite{DenHollander2000,Touchette2009,Dembo2010}. The existence of the rate function $I$ is referred to as the validity of a large deviation principle for the observable $C_n$ and can be seen as an extension of the weak law of large numbers as it provides information on the speed -- exponential in $n$ -- of convergence of $C_n$ to $c^*$.

In the context of Markov chains, there are several ways to compute the rate function $I$. It is known that, by means of spectral large deviation techniques (see, for instance, \cite{DeBacco2016,TsobgniNyawo2016,Whitelam2018,Coghi2019,Gutierrez2021,Whitelam2021}), one could calculate the scaled cumulant generating function (SCGF)
\begin{equation}
    \label{eq:SCGFdef}
    \Psi_N(s) \coloneqq \lim_{n \rightarrow \infty} \frac{1}{n} \ln \mathbb{E} \left[ e^{n s C_n} \right] \, ,
\end{equation}
where $s$ is the Lagrange (or tilting, in the large deviation jargon) parameter dual to $C_n=c$. The SCGF $\Psi$ represents the leading exponential behaviour of the moment generating function, associated with the observable in \eqref{eq:Cost}. To obtain the rate function $I$ in \eqref{eq:Rate}, it would then be enough to Legendre--Fenchel transform the SCGF, provided it be a differentiable function -- a result known as G\"{a}rtner--Ellis theorem \cite{DenHollander2000,Touchette2009,Dembo2010}. Although these methods serve well to the scope, variational techniques can also be employed and one may derive the rate function $I$ by solving a variational problem \cite{Maes2008,Barato2015,Chetrite2015a}. The advantage of employing variational methods is, at least, twofold. \deletetext{Indeed,} In case of non-analytically solvable problems, variational methods offer ways to bound the true rate function (see, for instance, \cite{Gingrich2016,Coghi2021}) and, at the same time, alternative numerical techniques -- inherited from the fields of optimization theory and PDEs -- are available to compute it \cite{Hoppenau2016}. 

In the case considered here, it is known that all the information on the fluctuations of one and two-point observables can be obtained by studying the pair empirical occupation measure
\begin{equation}
\label{eq:PairEmp}
L^{(2)}_n(i,j) = \frac{1}{n} \sum_{\ell=1}^n \delta_{X_{\ell},i}\delta_{X_{\ell+1},j} \hspace{1cm} \forall i,j \in \Gamma \, ,
\end{equation}
as the value of $C_n$ can be deduced via the formula
\begin{equation}
\label{eq:ContrForm}
C_n = \sum_{i,j=1}^N f(i,j) L_n^{(2)}(i,j) \, .
\end{equation}
Interestingly, the long-time behaviour of \eqref{eq:PairEmp}, denoted by $\rho = (\rho_{ij})_{i,j=1}^N$, can be interpreted as the amount of time that the Markov chain $X$ spends transiting from a state $i$ to a state $j$ of $\Gamma$ \cite{DenHollander2000,Touchette2009,Dembo2010}.

The pair empirical occupation measure of \eqref{eq:PairEmp} is known to satisfy a large deviation principle of the form
\begin{equation}
\label{eq:LDPPairEmp}
\mathbb{P}\left( L_n^{(2)} = \nu \right) = e^{-n H[\nu] + o(n)} \, ,
\end{equation}
with rate function
\begin{equation}
\label{eq:LevPair}
H[\nu] = \sum_{i,j} \nu_{ij} \ln \left( \frac{\nu_{ij}}{\mu_{i} \Pi_{ij}} \right) \, ,
\end{equation}
where $\nu = (\nu_{ij})_{i,j=1}^N$ belongs to the set of probability measures satisfying two constraints: the global balance on the state space, i.e., $\sum_j \nu_{ij} = \sum_j \nu_{ji}$, such that the sum of probability density currents flowing in and out of an arbitrary state $i$ is conserved, and the normalisation $\sum_{i,j} \nu_{ij} = 1$ (with $\sum_j \nu_{ij} = \mu_i$ the occupation measure). The rate function $H$ in \eqref{eq:LevPair} is known to be finite, continuous, and convex for densities $\nu$ that satisfy the global balance on the state space, featuring minimum and zero for $\nu = \rho$ \cite{DenHollander2000}. Here, it is interesting to notice that the rate function $I$ associated with $C_n$, can be obtained variationally by solving the following contraction\footnote{This term is commonly used in the large deviation theory jargon when referring to the solution of a variational problem passing from a higher-up level in the large deviation hierarchy to a lower one.} (minimisation) problem
\begin{equation}
\label{eq:ContrPairRate}
I(c) = \inf_{\substack{\nu: \\ c = \sum_{i,j} f(i,j) \nu_{ij}}} H[\nu] \, ,
\end{equation}
where the constraint appearing beneath the $\inf$ symbol is the formula \eqref{eq:ContrForm}, which selects $c$, the fluctuation of interest for the observable $C_n$ in \eqref{eq:Cost}.

The functional $H$ in \eqref{eq:LevPair} is thus a key ingredient for the variational study of fluctuations in discrete-time Markov chains and, as mentioned, it plays a pivotal role in statistical mechanics as many interesting dynamical observables arising in physics have the two-point form in \eqref{eq:Cost}.

The form in \eqref{eq:LevPair} has been derived with various methods. Among these, the exponential tilting procedure combined with the Radon--Nikodym change of measure \cite{DenHollander2000} (see \cite{Barato2015} for continuous-time processes) holds a leading position as it offers a simple and straightforward way to tackle the calculation, provided that the form of the rate function for the i.i.d.\ process (or any other useful process) is known. We will review and discuss this method in Section \ref{sec:ratehistory}. 

Although simple and well suited to large deviation estimates, the exponential tilting procedure does not allow for the calculation of $o(n)$-exponential sub-leading terms in the probability distribution of the pair empirical occupation measure \eqref{eq:PairEmp}. In the probability and applied statistics literature, however, exact combinatorial derivations that work at finite time can be found. These may lead to the evaluation of sub-leading order terms that, although not significant in the large deviation regime, would be important if one wanted to study transition regimes. The first combinatorial result goes back to \cite{Whittle1955}, later on reviewed in \cite{Billingsley1961}, and more recently recalled in \cite{CsiszaR1987}. Another graph-combinatorial derivation for the probability distribution of the pair empirical occupation measure was proposed in \cite{Dawson1957} and later on extended in \cite{Goodman1958}. \newtext{More recently, \cite{Polettini2015} provided an explicit---although not fully rigorous---expression of subleading terms in \eqref{eq:LDPPairEmp}, and constructed a Gauge theory for typical fluctuations of $C_n$ around its expected value. }

In the main Section \ref{sec:graphcombi} of our paper we use similar arguments to provide an alternative, exact expression for the moment generating function of the pair empirical occupation measure. We make use of notation and terminology that are more familiar to the theoretical physics audience, and show\newtext{---in line with previous literature \cite{Touchette2009}---}that our \deletetext{alternative} expression for the SCGF, akin to a Helmoltz (canonical) free energy, allows us to give a straightforward physical interpretation of all the terms and of the Lagrange multipliers that fix the necessary constraints\deletetext{,}\newtext{. Furthermore, we establish}  \deletetext{establishing }a direct link with spectral methods \newtext{and show an alternative variational formulation of the so-called driven process \cite{Jack2010,Chetrite2013,Chetrite2015,Chetrite2015a} (the Markov process responsible for the creation of fluctuations in the large-deviation regime)}. \newtext{In section \ref{sec:example} we show explicitly in a general two-state model the equivalence of our approach and the standard spectral techniques to compute the moment generating function at finite time $n$.}

\section{Pair empirical measure rate functional}
\label{sec:ratehistory}

In this Section we show how the rate functional $H$ in \eqref{eq:LevPair} can be derived via the exponential-tilting method. We start by writing the path-probability definition 
\begin{align}
\label{eq:FinProofMarkov0}
\mathbb{P} \left( L_n^{(2)} = \frac{T}{n} \right) \coloneqq \mathbb{P} \left( L_n^{(2)}(i,j) = \frac{t_{ij}}{n} \; \forall i,j \in \Gamma \right) &= \sum_{X_1,X_2,\ldots,X_{n+1}} \mathbb{P}(X_1,X_2,\cdots,X_{n+1}) \delta_{L_n^{(2)},T/n} \\
\label{eq:FinProofMarkov1}
&= \sum_{X_1,X_2,\dots,X_{n+1}} \mathbb{P}(X_1)\Pi_{X_1,X_2}\cdots \Pi_{X_{n},X_{n+1}} \delta_{L_n^{(2)},T/n} \, ,
\end{align}
where $t_{ij}$ represents the number of jumps that the Markov chain $X$ makes between nodes $i$ and $j$, and in \eqref{eq:FinProofMarkov1} we make use of the Markov property. We also notice in \eqref{eq:FinProofMarkov0} that we can interpret the set of $t_{ij}$s as the elements of a matrix $T$, which will be a central object in the rest of this work. 

We now introduce a new i.i.d.\ process $X' = \left( X'_{\ell} \right)_{\ell=1}^{n+1} = (X'_1,X'_2,\dots,X'_{n+1})$ based on the probability distribution $\zeta = \left( \zeta_i \right)_{i=1}^N$ on the state space and with its own pair empirical occupation measure that, with abuse of notation, have the same form of \eqref{eq:PairEmp}. A large deviation principle for the pair empirical measure of $X'$ is known to hold (see, for instance, Chapter 9 of \cite{Ellis1985} or Section II.2 of \cite{DenHollander2000}) with rate functional
\begin{equation}
\label{eq:LevPairIID}
H_{\text{i.i.d.}}[\nu] = \sum_{i,j} \nu_{ij} \ln  \left( \frac{\nu_{ij}}{\mu_{i}\zeta_{j}} \right) \, .
\end{equation} 
Consequently, we multiply and divide in the summation of \eqref{eq:FinProofMarkov1} by the path-probability $\mathbb{P}' ( L_n^{(2)} = T/n )$ of this i.i.d.\ process and then introduce an exponential function as follows
\begin{align}
\label{eq:FinProofMarkov2}
\mathbb{P} \left( L_n^{(2)} = \frac{T}{n} \right) &= \sum_{X_1,X_2,\dots,X_{n+1}} \mathbb{P}(X_1)\Pi_{X_1,X_2}\dots \Pi_{X_{n},X_{n+1}} \frac{\mathbb{P'}(X_1)\mathbb{P'}(X_2)\dots\mathbb{P'}(X_{n+1})}{\mathbb{P'}(X_1)\mathbb{P'}(X_2)\dots\mathbb{P'}(X_{n+1})} \delta_{L_n^{(2)},T/n}\\
\label{eq:FinProofMarkov3}
&= \sum_{X_1,X_2,\dots,X_{n+1}} \frac{\mathbb{P}(X_1)}{\mathbb{P'}(X_1)} e^{\sum_{\ell=1}^n \left[ \ln \Pi_{X_\ell,X_{\ell+1}} - \ln \zeta_{X_{\ell+1}} \right]}\mathbb{P'}(X_1)\mathbb{P'}(X_2)\dots\mathbb{P'}(X_{n+1}) \delta_{L_n^{(2)},T/n} \, .
\end{align}
The derivation continues by observing, in the exponential function, the equality
\begin{equation}
    \label{eq:Equality}
    \sum_{\ell=1}^n \left[ \ln \Pi_{X_\ell,X_{\ell+1}} - \ln \zeta_{X_{\ell+1}} \right] = n \sum_{i,j=1}^N L_n^{(2)}(i,j) \left[ \ln \Pi_{ij} - \ln \zeta_{j} \right] \, ,
\end{equation}
obtained by using the definition of the pair empirical measure \eqref{eq:LevPair}. Hence, we get
\begin{equation}
\label{eq:FinProofMarkov4}
\mathbb{P} \left( L_n^{(2)} = \frac{T}{n} \right) = \sum_{X_1,X_2,\dots,X_{n+1}} \frac{\mathbb{P}(X_1)}{\mathbb{P'}(X_1)} e^{n \sum_{i,j=1}^N L_n^{(2)}(i,j) \left[ \ln \Pi_{ij} - \ln \zeta_{j} \right]}\mathbb{P'}(X_1)\mathbb{P'}(X_2)\dots\mathbb{P'}(X_{n+1}) \delta_{L_n^{(2)},T/n} \, .
\end{equation}
Eventually, by taking minus the logarithm of the probability $\mathbb{P}$, dividing by $n$, and taking the limit $n \rightarrow \infty$ we get
\begin{align}
\nonumber
- \lim_{n \rightarrow \infty} \frac{1}{n} \ln \mathbb{P} \left( L_n^{(2)} = \nu \right) &= \sum_{i,j=1}^N \nu_{ij} \ln \frac{\zeta_{j}}{\Pi_{ij}} + \lim_{n \rightarrow \infty} \frac{1}{n} \ln \left( \sum_{X_1,X_2,\dots,X_{n+1}} \frac{\mathbb{P}(X_1)}{\mathbb{P'}(X_1)} \mathbb{P'}(X_1)\mathbb{P'}(X_2)\dots\mathbb{P'}(X_{n+1}) \delta_{L_n^{(2)},\nu} \right) \\
\label{eq:FinProofMarkov5}
&= \sum_{i,j=1}^N \nu_{ij} \left( \ln \frac{\zeta_{j}}{\Pi_{ij}} - \ln \left( \frac{\nu_{ij}}{\mu_{i} \zeta_{j}} \right) \right) \, ,
\end{align}
where the matrix $\nu$ is defined as
\begin{equation}\label{eq:nuT}
    \nu = \frac{T}{n} \, .
\end{equation}

In the derivation of \eqref{eq:FinProofMarkov5}, we make use of the fact that $L_n^{(2)} \rightarrow \nu$, and also that for the probability $\mathbb{P}'(L_n^{(2)}=\nu)$ a large deviation principle holds with rate functional \eqref{eq:LevPairIID}. The last formula obtained in \eqref{eq:FinProofMarkov5} is exactly \eqref{eq:LevPair}. We remark that it is only because of the long-time limit that we can get rid of the boundary term $\mathbb{P}(X_1)/\mathbb{P}'(X_1)$ in \eqref{eq:FinProofMarkov5} and thus get the form of the rate functional for the pair empirical occupation measure of the Markov process $X$. We also notice that, although extremely useful, the use of an i.i.d.\ process with its pair empirical rate functional is not strictly necessary for the purpose of the proof. Indeed, if the asymptotics of the pair empirical probability of another process were known and easy to handle, we could have tilted the path probability measure of the Markov process in \eqref{eq:FinProofMarkov2} with respect to it and we would have obtained the same result. For further details on this and on how to best use the tilting method we refer to \cite{Chetrite2015}.

The derivation presented in this Section makes use of methods that are well known in the large deviation community. Nevertheless, for a more rigorous proof of the large deviation principle for the pair empirical measure \eqref{eq:PairEmp} having rate functional \eqref{eq:LevPair} -- which focuses on lower and upper bounds over closed and open sets -- we refer the reader to \cite{Ellis1985,DenHollander2000,Dembo2010}.

\deletetext{For what matters here, we notice that} The derivation presented in this Section takes into consideration only leading order terms in $n$ and, furthermore, \deletetext{may} lacks some physical interpretations of the form of the rate functional $\eqref{eq:LevPair}$. \deletetext{To make up for these shortcomings, in the next Section, we provide a graph-combinatorial derivation \cite{Dawson1957,Goodman1958} that allows us to get the exact form of the moment generating function of the pair empirical occupation measure in \eqref{eq:PairEmp}.} \newtext{The finite $n$ behaviour, captured by subleading terms in \eqref{eq:LDPPairEmp}, is in general much harder to study than the large deviations regime. For the continuous-time setting, in \cite{Causer2022} the authors use matrix product states to study finite-time large fluctuations of one-dimensional lattice models. For discrete Markov Chains, estimates and bounds for subleading terms in \eqref{eq:LDPPairEmp} are known in the literature \cite{kontoyiannis2003spectral,kontoyiannis2005large}, and derived by using spectral methods. In \cite{Polettini2015} the author proposes a characterisation of subleading terms using graph-combinatorial arguments. Using a similar approach as \cite{Polettini2015}, we provide an exact formula for the moment generating function valid for any finite $n$, a first step towards an alternative derivation of the subleading terms in \eqref{eq:LDPPairEmp}.}

\section{Graph-combinatorial approach}
\label{sec:graphcombi}

In this Section, we present an alternative derivation of the rate functional associated with the pair empirical occupation measure in \eqref{eq:PairEmp}. The proposed derivation moves the focus from the probability distribution $\mathbb{P}$ and rate functional $H$ in \eqref{eq:PairEmp} to the moment generating function $Z_{N,n}$ and SCGF
\begin{equation}
    \label{eq:SCGF}
    \lambda_N[s] \coloneqq \lim_{n \rightarrow \infty} \frac{1}{n} \ln Z_{N,n}[s] = \lim_{n \rightarrow \infty} \frac{1}{n} \ln \mathbb{E} \left[ e^{n s \cdot L_n^{(2)}} \right] \,
\end{equation}
where, with abuse of notation with respect to \eqref{eq:SCGFdef}, $s=\left( s_{ij} \right)_{i,j=1}^N$ is now a set of Lagrange parameters.
This paradigm shift is equivalent to a change of ensemble in statistical mechanics \cite{Chetrite2015}. Instead of working with the probability distribution $\mathbb{P}( L_n^{(2)}=T/n )$ at a fixed $t$, we introduce Lagrange parameters $s_{ij}$s that fix the $t_{ij}$s only on average, and thus work with a moment generating function. The equilibrium statistical mechanics analogue would be a change from the microcanonical ensemble, where the energy is fixed, to the canonical ensemble, where only the average energy is fixed by the Lagrange parameter $\beta$, the inverse temperature.

In this canonical framework, thanks to Markovianity and ergodicity, it is known \cite{Touchette2009,Dembo2010} that we can map the large deviation problem to a spectral one. This is because the SCGF can be calculated as the logarithm of the dominant eigenvalue of the so-called tilted matrix $\Pi_s = \left( \Pi_s \right)_{ij} \forall i,j \in \Gamma$, which has the form
\begin{equation}
    \label{eq:TiltMatr}
    \left( \Pi_s \right)_{ij} = \Pi_{ij} e^{s_{ij}} .
\end{equation}

Noticeably, thanks to a graph-combinatorial mapping \cite{Dawson1957,Goodman1958}, we can derive an exact expression for the moment generating function $Z_{N,n}$ at finite $N$ and $n$. In principle, the exact form $Z_{N,n}$ allows one to evaluate sub-leading terms (in $n$) that cannot be calculated within a purely large deviation approach as that of Section \ref{sec:ratehistory}. Historically, graph-combinatorial arguments similar to those used in this work have been proposed for cyclic Markov chains by Dawson and Good in \cite{Dawson1957}, and later on extended for general Markovian paths by Goodman in \cite{Goodman1958}. The derivation that follows explains in the details, with a theoretical-physics approach, a similar graph-combinatorial calculation but moves the focus onto the moment generating function $Z_{N,n}$ of the pair empirical occupation measure. This allows us to naturally give a physical interpretation of the interaction and entropic terms in the SCGF $\lambda_N$ \eqref{eq:SCGF}.

\subsection{An alternative expression for the moment generating function}

The graph-combinatorial approach is based on the representation of the state space connectivity as a graph $\mathbf{G}$ with associated adjacency matrix $A$ -- see Fig.\ \ref{fig1}(a). This has elements $A_{ij} = 1$ if state $j$ can directly be reached from $i$, and $0$ otherwise. In this context, we will refer to states also as nodes or vertices. The transition matrix $\Pi$ of the Markov chain $X$, in turn, embeds in its elements the connectivity of the state space as $\Pi_{ij} = A_{ij} p_{ij}$ with $p_{ij}$ the jump probability between $i$ and $j$.

The moment generating function of the probability $\mathbb{P} ( L_n^{(2)} = T/n)$ is 
\begin{equation}\label{eq:moment_trajectories}
    Z_{N,n}(s)  =  \sum_{X_1, \dots, X_{n+1}}   \mathbb{P}(X_1) \prod_{\ell=1}^{n} \Pi_{X_{\ell}, X_{\ell+1}} e^{\sum_{ij} s_{ij} \delta_{X_{\ell},i} \delta_{X_{\ell +1},j}}  \ ,
\end{equation}
where $\mathbb{P}(X_1)$ indicates the probability distribution of our process at initial time $n=1$ and $s = (s_{ij})_{i,j=1}^N$ indicates the set of tilting parameters. 

The specific form of the distribution $\mathbb{P}(X_1)$ will play a role for finite time behaviour or sub-leading asymptotics, but it will not matter in the large deviation regime -- it only amounts to a boundary term -- provided that the graph $\mathbf{G}$ is connected. For convenience, we choose $\mathbb{P}(X_1) = \delta_{X_1,1}$, viz.\ the starting node is fixed to be node $1$. 

The core idea of our work is to perform a change of variables: we transform the sum over all states $X_\ell$, for $\ell \in \{1, \dots, n+1\}$, to a sum over variables $t_{ij} \in \{0, \dots, n\}$, for nodes $i,j \in \mathbf{G}$. The new variable $t_{ij}$, as in the previous Section \ref{sec:ratehistory}, is the number of times the Markov chain jumps from state $i$ to state $j$, in particular $t_{ij}$ can be different from zero only if there is an edge in $\mathbf{G}$ between nodes $i$ and $j$. 

For a matrix $T$ to represent the number of jumps of a chain of states $(X_1, X_2, \dots, X_{n+1})$ the following constraints must be satisfied: (i) the total number of jumps is equal to the total length of the chain minus one, $\sum_{ij} t_{ij} = n$, (ii) all jumps can be temporally arranged like domino tiles $(1,X_2), (X_2, X_3), \dots, (X_{n}, X_{n+1})$ reflecting the fact that if at time $\ell$ the Markov chain jumps to state $i$, then at time $\ell + 1$ it has to start from state $i$. Constraints (i) and (ii) do not make the change of variables one to one -- there can be many instances of the Markov chain that correspond to the same set of $t_{ij}$s. In fact, the variables $t_{ij}$ do not carry any information regarding the temporal order of the jumps. In other words, given an instance of $T$ we have to count in how many ways we can order the jumps as $(1, X_2), (X_2, X_3), \dots, (X_n, X_{n+1})$ so that $\sum_{\ell=1}^n \delta_{X_{\ell},i}\delta_{X_{\ell+1},j} = t_{ij} $ and  $\{X_1, X_2, \dots, X_{n+1}  \}$ realises a walk in $\mathbf{G}$: we call this number $\Theta_T$. Hence, we can express $Z_{N,n}$ as
\begin{equation}\label{eq:changeofvariablest}
    Z_{N,n}(s) = \sum_{t_{11} = 0}^{n} \dots \sum_{t_{ij} = 0}^n \dots \sum_{t_{NN} = 0}^n \delta_{\sum_{ij} t_{ij}, n}  \ \Theta_T  \prod_{i,j} \left( \Pi_{ij} ^{t_{ij}}  e^{ s_{ij} t_{ij} } \right) \ .
\end{equation}

\begin{figure}[htb]
\centering
\includegraphics[width=\textwidth]{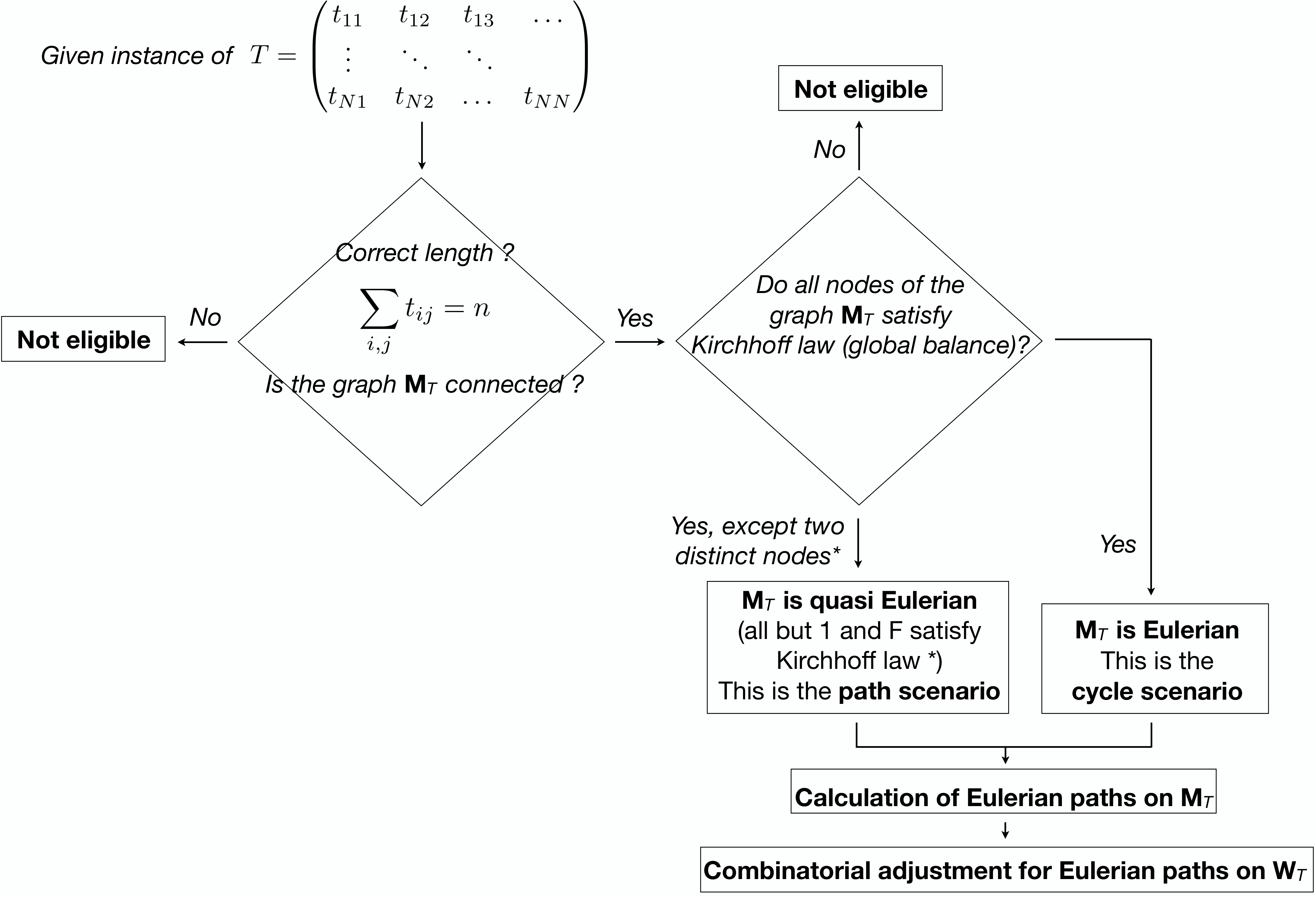}
\caption{Flowchart that summarizes the computation of $\Theta_T$. From top-left, we start from an instance of the matrix of jumps $T$. We first check if the total number of jumps is $n$ and if the multi-graph $\mathbf{M}_T$ associated with $T$ is connected. Then, we proceed by checking if Kirchhoff law is satisfied. There are two possible positive scenarios: i) it is satisfied in every node -- \textit{cycle} scenario; ii) it is satisfied in every node except node $1$ and another node which we call $F$: in the former node there is one more outgoing link, while in the latter there is one more incoming link -- \textit{path} scenario. In either of these cases, we can relate $\Theta_T$ to the number of T-eulerian cycles (or paths) in  $\mathbf{W}_T$, which can be computed knowing the number of eulerian cycles (or paths) in $\mathbf{M}_T$.}
\label{fig:rat}
\end{figure}

We now face the problem of computing $\Theta_T$. We notice that for many instances of $T$, this number is simply zero: this is because the aforementioned domino-like constraint (ii) imposes stringent conditions on the form of $T$. First of all, the set of edges $(i,j)$, for which $t_{ij}>0$, together with the union of all their extremes $i$ and $j$ must form a connected graph. This is because the Markov chain starting from a node $i$ can only hop to neighbours of $i$ according to the connectivity of $G$.
%This is because the Markov chain cannot disappear from a node and re-appear in another one without us noticing \fra{Perhaps we can write this better} -- the variables $T$ account for every jump of the Markov chain.
Mathematically, this condition is equivalent to requiring that the dimension of the kernel of the Laplacian $L = D_{\mathrm{in}} - T$ is $1$ \cite{Newman2010}, where $D_{\mathrm{in}}$ is a diagonal matrix with elements $(D_{\mathrm{in}})_{ii} = \sum_{j} t_{ji} $. Second, the number of times a Markov chain jumps towards a state $i$ have to be related to the number of jumps starting from that state $i$, a phenomenon analogous to the Kirchhoff law in electric circuits that encodes the global balance of the dynamics. We hereby distinguish two possible scenarios in which these conditions on $T$ are satisfied. In the first one, for every state the incoming flux and outgoing flux are equal, that is $\sum_{j} t_{ij} = \sum_{j} t_{ji}$: this situation corresponds to a Markov chain starting and ending in the same node, and we will refer to this as the \textit{cycle} scenario. In the second one, for all but two states the incoming and outgoing fluxes are equal. The two special states are the initial, that we set to $1$ choosing $\mathbb{P}(X_1) = \delta_{X_1,1}$, and the final, $F$, for which one must have $\sum_j t_{1j} = 1 + \sum_j t_{j1}$ and $1+\sum_j t_{Fj} =  \sum_j t_{jF}$: we will refer to this as the \textit{path} scenario. This leads to a natural way to express

\begin{align}
    & \Theta_T = \Theta_T^{\text{path}} \sum_{F \neq 1}\left(\prod\limits_{i \neq 1, F}^N \delta_{\sum_{j=1}^Nt_{ij},\  \sum_{j=1}^N t_{ji}}\right) \delta_{\sum_{j=1}^Nt_{1j}+1,\  \sum_{j=1}^N t_{j1}}\delta_{\sum_{j=1}^Nt_{Fj},\  \sum_{j=1}^N t_{jF}+1}  + \nonumber \\
    & + \Theta_T^{\text{cycle}} \prod\limits_{i=1}^N \delta_{\sum_{j=1}^Nt_{ij},\  \sum_{j=1}^N t_{ji}} \left(1 - \delta_{\sum_{j = 1}^N t_{1j}, 0} \right)  \ ,
\end{align}

where $\Theta_T^{\text{path}}$ ($\Theta_T^{\text{cycle}}$) is the number of distinct permutations of the set of $t_{ij}$s in the path (cycle) scenario which give a realisation of a walk in $\mathbf{G}$ and the deltas enforce Kirchhoff law. \newtext{The factor $1 - \delta_{\sum_{j = 1}^N t_{1j}, 0}$ ensures that the cycle will pass at least once from node $1$: this condition is required because the starting node is node $1$.} We will show that it is not necessary to enforce connectedness explicitly because the expressions for $\Theta_T^{\text{path}}$ and $\Theta_T^{\text{cycle}}$ are automatically zero when $T$ is not connected.

We now note that we can interpret the matrix $T$ as the adjacency matrix of a directed multi-graph $\mathbf{M}_T$ with $t_{ij}$ directed links, having unitary weight, between nodes $i$ and node $j$ -- see Fig.\ \ref{fig1}(b). A directed multi-graph is a collection of nodes and directed links, in which multiple links between two nodes are permitted. We refer to the collection of links between two nodes as a multi-link. 
As a preliminary step in the computation of $\Theta_T^{\text{path}}$  ($\Theta_T^{\text{cycle}}$), we  consider a related combinatorial problem, that is counting how many paths there are on $\mathbf{M}_T$ that start in $1$ and end in $F$ (cycles that start in $1$) and pass through every link exactly once. We can interpret this as the number of \textit{non-distinct} ways we can arrange the jumps like domino tiles that respect the matrix of jumps $T$. This number overestimates $\Theta_T^{\text{path}}$  (respectively $\Theta_T^{\text{cycle}}$). To see this, we can consider a multi-link in $\mathbf{M}_T$ having at least two links $l_1$ and $l_2$. Given a path (or cycle) that passes through every link in $\mathbf{M}_T$, we can, for instance, generate another distinct one by swapping the order in which we visit $l_1$ and $l_2$. This new path (cycle) will not contribute to $\Theta_T^{\text{path}}$  ($\Theta_T^{\text{cycle}}$), as the time-ordered jumps $(1,X_2), (X_2,X_3), \dots, (X_n, X_{n+1})$ are unaffected by the swap.
Nonetheless, this calculation is a useful starting point as we can compute this number using results available in the literature \cite{T.Aardenne-Ehrenfest1951}, and we will show how to correct this overcounting later on.

\begin{figure}[htp]
\centering
\includegraphics[width=.3\textwidth]{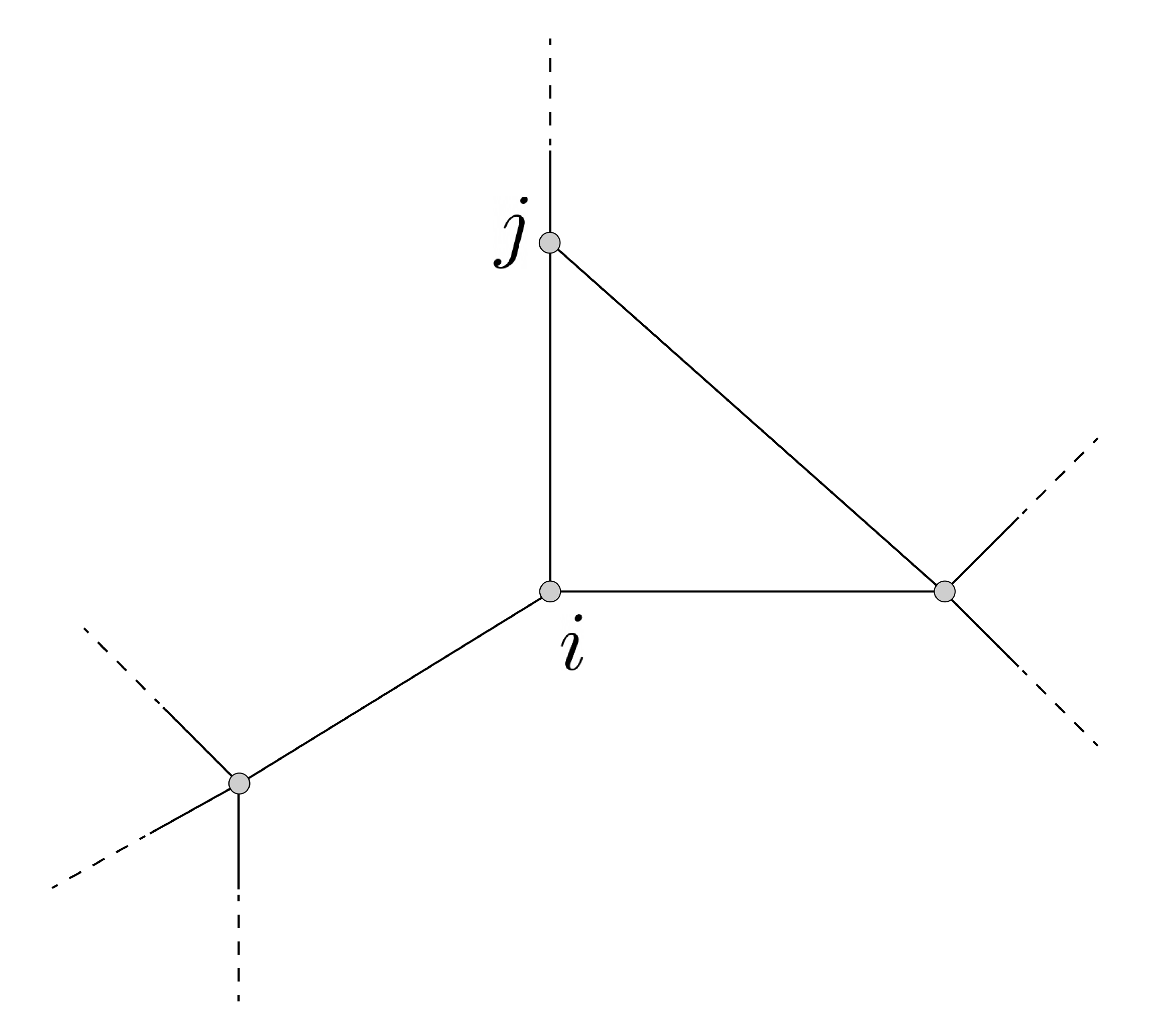}\hfill
\includegraphics[width=.3\textwidth]{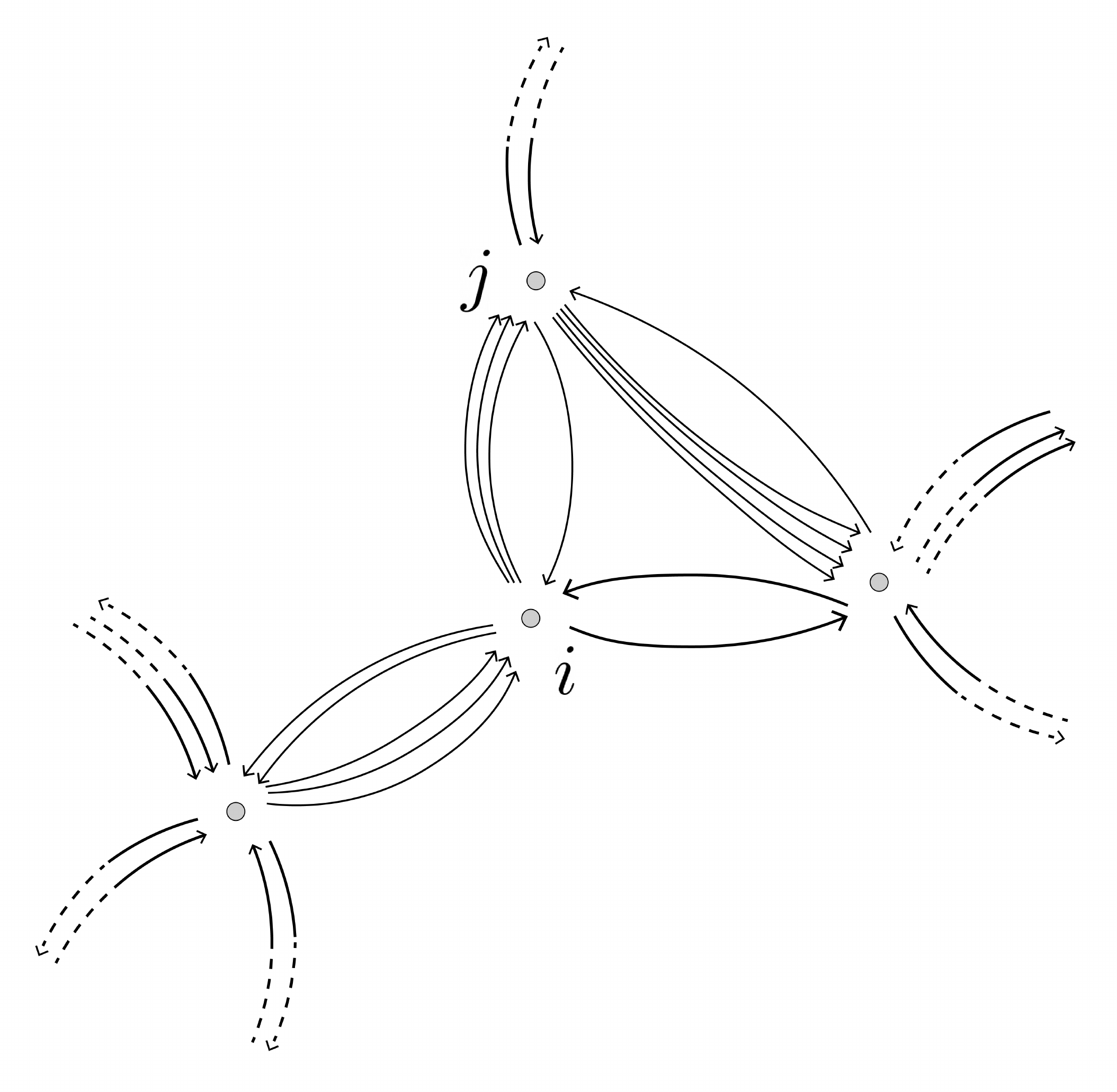}\hfill
\includegraphics[width=.3\textwidth]{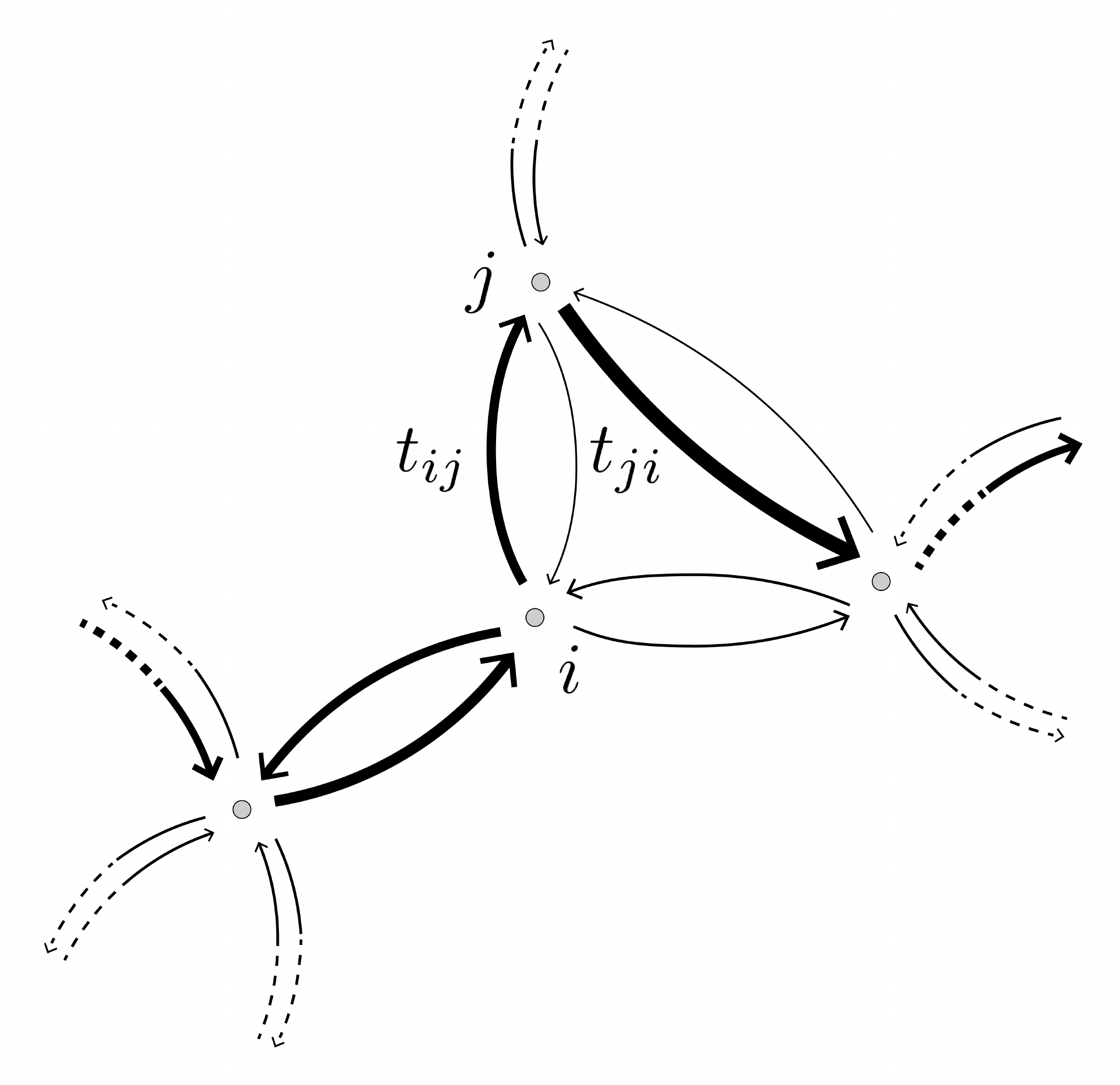}
\put(-470,110){(a)}
\put(-310,110){(b)}
\put(-140,110){(c)}
\caption{(a) State-space connectivity $\mathbf{G}$, un-directed and un-weighted, with adjacency matrix $A$; (b) directed un-weighted multi-graph $\mathbf{M}_T$, with adjacency matrix $T$, a multi-link from $i$ to $j$ is composed by $t_{ij}$ links; (c) directed weighted graph $\mathbf{W}_T$, with adjacency matrix $T$, the boldness of links is proportional to the integer weights $t_{ij}$. }
\label{fig1}
\end{figure}

\subsection{Computation of $\Theta_T^{\text{path}}$ and $\Theta_T^{\text{cycle}}$}

So far we have described the key steps that underlie our approach, which are also summarised in the flowchart in Fig.\ \ref{fig:rat}. In the following instead we will provide the details of the calculation. For this reason, some definitions will be useful and we collect them in this paragraph. An eulerian multi-graph is a multi-graph for which, at every node $i$, in-degree and out-degree are the same, viz.\ $k_i^{in} = k_i^{out}$. Noticing that in $\mathbf{M}_T$ we have $k_i^{in} = \sum_j t_{ji}$ and $k_i^{out} = \sum_j t_{ij}$, it follows from Kirchhoff law that $\mathbf{M}_T$ is either an eulerian multi-graph (for the cycle scenario) or close to an eulerian multi-graph (for the path scenario), in the sense that only the initial and final nodes do not satisfy $k_{in} = k_{out}$. An eulerian cycle (path) on a multi-graph $\mathbf{M}_T$, as already mentioned in Section \ref{sec:ratehistory}, is a cycle (path) that passes through every link exactly once. We denote the number of eulerian cycles (paths) with $ec(\mathbf{M}_T)$ ($ep(\mathbf{M}_T)$). Furthermore, we will indicate by $ec(\mathbf{M}_T|\chi)$ ($ep(\mathbf{M}_T|\chi)$) the number of eulerian cycles (paths) given some specified condition $\chi$, that in our case will be a combination of the starting node $1$, the final node $F$, the starting edge $e_{1}$ and the final edge $e_F$. 

In the literature on the topic, a result is known for the number $ec(\mathbf{M}_T|e_1)$ of eulerian cycles of an eulerian multi-graph $\mathbf{M}_T$ with a fixed starting edge. This goes by the name of BEST theorem \cite{T.Aardenne-Ehrenfest1951, Tutte1941, Farrell2015} and reads 
\begin{equation}
    \label{eq:Best}
    ec(\mathbf{M}_T|e_1) =  \Omega_w(\mathbf{M}_T) \prod_{i=1}^N (k^{in}_i - 1)!,
\end{equation}
where $\Omega_w(\mathbf{M}_T)$ is the number of arborescences, i.e., spanning trees rooted in a node $w$ such that there exists a unique path from every vertex of $\mathbf{M}_T$ to $w$. We note that $\Omega_w$ does not depend on the choice of root $w$ when $\mathbf{M}_T$ is an eulerian multi-graph, so that $\Omega_w(\mathbf{M}_T) = \Omega(\mathbf{M}_T)$ \cite{Tutte1941, Rubey2000}. Similarly, the r.h.s.\ of \eqref{eq:Best} does not show any explicit dependence on the starting edge $e_1$ because of the inherent symmetry in $\mathbf{M}_T$. An explicit expression for $\Omega(\mathbf{M}_T)$ is given by 
\begin{equation}\label{eq:matrixtreetheorem}
    \Omega(\mathbf{M}_T) = \det (L_w) \ ,
\end{equation}
where $\text{det}$ is the determinant operator and $L_w$ is a submatrix of the Laplacian of the eulerian multi-graph $\mathbf{M}_T$ obtained by removing (any) $w$-th row and column, a result known in the literature as Tutte's theorem or Matrix tree theorem.

In the following, we first consider the \textit{path scenario}. In this case, the multi-graph $\mathbf{M}_T$ is not eulerian, but we can make it so simply by adding a link $e_F$ from $F$ to $1$. We refer to this modified graph as $\tilde{\mathbf{M}}_T$. Using BEST theorem we have
\begin{equation}
ec(\tilde{\mathbf{M}}_T|e_1) = \Omega(\tilde{\mathbf{M}}_T) \prod_{i\neq 1}^N \left( \sum_{j=1}^Nt_{ji}- 1 \right)! \left(\sum_{j=1}^N t_{j1} \right)\ ,
\end{equation}
where we use the fact that the in-degree of node $1$ is $\sum_{j=1}^N t_{j1} + 1$ in $\tilde{\mathbf{M}}_T$. The number of eulerian cycles starting from node $1$ is related to $ec(\tilde{\mathbf{M}}_T|e_1)$ by $ec(\tilde{\mathbf{M}}_T| 1) = ec(\tilde{\mathbf{M}}_T|e_1) \left(\sum_j t_{j1} + 1 \right)$. Furthermore, $ep(\mathbf{M}_T|1,F)$ is equal to the number of eulerian cycles in $\tilde{\mathbf{M}}_T$ starting in $1$ and ending with the link we added to construct it, viz.\ $ec(\tilde{\mathbf{M}}_T|1,e_F)$. This number can be computed by considering an eulerian cycle in $\mathbf{M}_T$ as a collection of loops passing through node $1$. The number of these loops is given by the in-degree of node $1$, so that we have $ep(\mathbf{M}_T|1,F) = ec(\tilde{\mathbf{M}}_T| 1)/(\sum_j t_{j1} + 1)$. All these considerations put together give
\begin{equation}
    ep(\mathbf{M}_T|1,F) = \Omega_1(\mathbf{M}_T) \prod_{i\neq 1}^N \left( \sum_{j=1}^Nt_{ji}- 1 \right)! \left(\sum_{j=1}^N t_{j1} \right)\
\end{equation}

where we used $\Omega(\tilde{\mathbf{M}}_T) = \text{det}(L_1) =  \Omega_1(\mathbf{M}_T)$ with $L_1$ the cofactor of the graph Laplacian $L$ obtained by removing the first row and column. We note that while $\Omega(\tilde{\mathbf{M}}_T)$ does not depend on the choice of the root since $\tilde{\mathbf{M}}_T$ is eulerian, $\Omega_1(\mathbf{M}_T)$ does, because $\mathbf{M}_T$ is not eulerian. 

We now consider the \textit{cycle scenario}. In this case, since $\mathbf{M}_T$ is already an eulerian graph, we can readily express $ec(\mathbf{M}_T,1)$ as
\begin{equation}
    ec(\mathbf{M}_T|1) = ec(\mathbf{M}_T|e_1) \sum_{j=1}^N t_{j1} = \Omega(\mathbf{M}_T) \prod_{i\neq 1}^N \left(\sum_{j=1}^Nt_{ji}- 1\right)! \left(\sum_{j=1}^N t_{j1} \right)\ .
\end{equation}
As previously argued, $ep(\mathbf{M}_T|1,F)$ ($ ec(\mathbf{M}_T|1)$) overestimates $\Theta_T^{\text{path}}$ ($\Theta_T^{\text{cycle}}$). To correct this, one must consider all the links belonging to a given multi-link as totally equivalent. This boils down to considering a weighted graph $\mathbf{W}_T$ (see \ref{fig1}(c)) in place of the multi-graph $\mathbf{M}_T$. The weighted graph $\mathbf{W}_T$ has adjacency matrix $T$ and directed links (e.g., between nodes $i$ and $j$) obtained by merging all the multi-links (between $i$ and $j$) in $\mathbf{M}_T$ together. In analogy with \cite{Farrell2015}, we define the notion of $T$-eulerian cycle (path) as a cycle (path) that passes through every link $(i,j)$ a number $t_{ij}$ of times. With an abuse of notation, we denote the number of $T$-eulerian cycles (paths) by $ec(\mathbf{W}_T|\chi)$ ($ep(\mathbf{W}_T|\chi)$), as it will be clear by the graph we are considering whether we are referring to eulerian or $T$-eulerian cycles (paths). Crucially, in the cycle scenario $\Theta_T^{\text{cycle}}$ is equal to the number of $T$-eulerian cycles starting from node $1$, i.e., $ec(\mathbf{W}_T|1)$ in $\mathbf{W}_T$, whereas in the path scenario $\Theta_T^{\text{path}}$ is equal to the number of $T$-eulerian paths from $1$ to $F$, i.e., $ep(\mathbf{W}_T| 1, F)$ in $\mathbf{W}_T$. The combinatorial factor connecting $ep(\mathbf{W}_T| 1, F) $ ($ec(\mathbf{W}_T| 1)$ ) to $ep(\mathbf{M}_T| 1, F)$ ($ec(\mathbf{M}_T| 1)$) is simply the number of permutations of links in a multi-link for every multi-link in $\mathbf{M}_T$
\begin{align} \label{eq:connectionGTMT}
\begin{split}
    &ep(\mathbf{M}_T| 1, F) = ep(\mathbf{W}_T| 1, F) \prod_{i,j=1}^N t_{ij}!\\ 
    &ec(\mathbf{M}_T| 1) = ec(\mathbf{W}_T| 1) \prod_{i,j=1}^N t_{ij}! \ .
\end{split}
\end{align}

This allows us to write explicit expressions for $\Theta_T^{\text{path}}$ and $\Theta_T^{\text{cycle}}$
\begin{equation}\label{eq:ThetaFin}
    \Theta_T^{\text{path}} = \Theta_T^{\text{cycle}} =  \det (L_1) \prod_{i=1}^{N} \frac{( \sum_{j=1}^N t _{ji}-1)!}{ \prod_{j=1}^N t_{ij}!}  \sum_{k=1}^N t_{k1}  \ ,
\end{equation}
where we recall that $L_1$ is -- in both the cycle and the path scenarios --  the submatrix of the graph Laplacian $L$ obtained by removing the first row and column. We remark that, although the expressions for $\Theta_T^{\text{path}} $ and $\Theta_T^{\text{cycle}} $ in \eqref{eq:ThetaFin} are formally the same, the variable $T$ is of different nature in the path and cycle scenario as it satisfies different sets of constraints. With this expression, we can write the moment generating function explicitly as
\begin{align}\label{eq:permutationperformed}
\begin{split}
    Z_{N,n}(s) = &\sum_{t_{11} = 0}^{n} \dots \sum_{t_{ij} = 0}^n \dots \sum_{t_{NN} = 0}^n \delta_{\sum_{ij} t_{ij}, n} \  \prod_{i,j} \left(\Pi_{ij}^{t_{ij}} e^{ s_{ij} t_{ij}} \right) \det(L_1)  \sum_{j=1}^N t_{j1} \prod_{i=1}^{N} \frac{( \sum_{j=1}^N t _{ji}-1)!}{ \prod_{j=1}^N t_{ij}!}\\
   &\Bigg (  \sum\limits_{F \neq 1} \left(\prod\limits_{i \neq 1, F}^N \delta_{\sum_{j=1}^Nt_{ij},\  \sum_{j=1}^N t_{ji}} \right) \delta_{\sum_{j=1}^Nt_{1j}+1,\  \sum_{j=1}^N t_{j1}}\delta_{\sum_{j=1}^Nt_{Fj},\  \sum_{j=1}^N t_{jF}+1}  + \\ 
   & \left(\prod\limits_{i=1}^N  \delta_{\sum_{j=1}^Nt_{ij},\  \sum_{j=1}^N t_{ji}}\right) \left(1 - \delta_{\sum_j t_{1j},0} \right) \Bigg)   \ .
\end{split}
\end{align}
We note that the factor $\det(L_1)$ kills configurations of $T$ that have the null-space dimension of the Laplacian greater than $1$. This ensures that we only consider graphs $\mathbf{M}_T$ -- equivalently, $\mathbf{W}_T$ -- that are connected, as it is known in the literature that the dimension of the null-space of the graph Laplacian is the number of connected components of a graph \cite{Newman2010}. \newtext{ Remarkably, in equation \eqref{eq:permutationperformed} the contributions for paths and cycles are split, giving an interesting physical perspective. In general, this difference is more pronounced when $n$ is small, in particular when the walker has not explored the full state space. In the limit of large $n$, contributions relative to paths and cycles are comparable and share the same asymptotics, as we show in the next Section.

Compared to the spectral method to compute the moment generating function \cite{Touchette2009}, which requires the computation of all eigenvalues and eigenvectors of an $N \times N$ matrix, our formula is computationally favourable when $n$ is small and $N$ is large. If $n$ is large, instead, the spectral method is numerically more efficient.}

\subsection{Long-time asymptotics}

Expression (\ref{eq:permutationperformed}) is valid for every finite $n$, and can be used to derive the large $n$ limit and, in principle, finite $n$ corrections. In the following, we focus on the large deviation regime, which corresponds to taking $n$ to be much greater than the longest relaxation time of the system $\tau(N)$,  $n \gg \tau(N)$. In this limit it is useful to rescale time-additive variables with $n$ as in \eqref{eq:nuT}, as we can approximate the sums over $t_{11}, \dots, t_{NN}$ with integrals
\begin{equation}\label{eq:SumtoInt}
    \frac{1}{n^{|E_{\mathbf{W}_T}|}}\sum_{t_{11}} \dots \sum_{t_{NN}} \ \to \; \prod_{i,j}\int_0^1 \! \mathrm{d}\nu_{ij}\ , 
\end{equation}
where $|E_{\mathbf{W}_T}|$ is the number of directed edges in the weighted graph $\mathbf{W}_T$ and $\nu_{ij}$s are defined as in \eqref{eq:nuT}. In the r.h.s.\ of \eqref{eq:SumtoInt} and in the following, by $\sum_{ij}$ and $\prod_{ij}$ we mean sums and products over $(i,j)$ such that $(i,j)$ is a directed link in $\mathbf{W}_T$. To leading order in $n$ we obtain the following asymptotic expressions
\begin{align}
    &\prod_{i,j} \left(\Pi_{i j}^{t_{ij}} e^{ s_{ij} t_{ij}} \right) \ \to \; e^{n \sum_{i=1}^N\sum_{j=1}^N \left(s_{ij} + \log \Pi_{ij} \right)\  \nu_{ij}}\\ 
    &\left( \sum_{j=1}^N t_{j1} \right) \left(\prod_{i=1}^{N} \frac{( \sum_{j=1}^N t_{ji}-1)!}{ \prod_{j=1}^N t_{ij}!} \right)\ \to  \; e^{n \sum_{i=1}^N\sum_{j=1}^N \nu_{ij} (\log(\sum_{k=1}^N \nu_{ik}) - \log(\nu_{ij}))}\\ 
    &\delta_{\sum_{ij} t_{ij}, n} \  \to \; \delta_{\sum_{ij} \nu_{ij}, 1} \\
    &n^{|E_{\mathbf{W}_T}|} \ \to \; e^{|E_{\mathbf{W}_T}| \ \log n}  \ .
\end{align}
The Kirchhoff constraints tend to the same form for large $n$, giving explicitly
\begin{align}
&\sum_{F \neq 1} \left( \prod_{i \neq 1,F} \delta_{\sum_{j=1}^N t_{ji},\sum_{j=1}^N t_{ij}}\right) \delta_{\sum_{j=1}^N t_{j1}+1,\sum_{j=1}^N t_{1j}} \delta_{\sum_{j=1}^N t_{jF},1+\sum_{j=1}^N t_{Fj}} \ \to \; (N-1)\prod_{i=1}^N \delta_{\sum_{j=1}^N \nu_{ji},\sum_{j=1}^N \nu_{ij}} \\
& \prod_{i=1}^N \delta_{\sum_{j=1}^N t_{ji},\sum_{j=1}^N t_{ij}} \to \; \prod_{i=1}^N \delta_{\sum_{j=1}^N \nu_{ji},\sum_{j=1}^N \nu_{ij}} \ .
\end{align}
We also notice that 
\begin{equation}
    \text{det} (L_1) = n^{N-1} \det \left( \frac{L_1}{n} \right) =  e^{(N-1) \log n + \text{Tr}  \left[\log \left(\frac{L_1}{n}\right) \right]}\  \to \; e^{(N-1) \log n} \ ,
\end{equation}
where we use the fact that the determinant is multi-linear in the rows and that each element in $L_1$ is proportional to $n$ by construction, so that $\text{Tr}  \left[\log \left(\frac{L_1}{n}\right) \right]$ is finite for large $n$. Remarkably $\text{det}(L_1)$ becomes sub-leading in the large $n$ limit, while it may be an interesting term to study the finite-time transient behaviour of the Markov chain.
\newtext{Finally, the term $\delta_{\sum_j t_{1j},0}$ present in the factor $1 - \delta_{\sum_j t_{1j},0}$ becomes negligible for large $n$. }

\deletetext{Eventually}\newtext{Putting all together}, we obtain to exponential leading order in $n$
\begin{equation}\label{eq:largenmgf}
    Z_{N,n}(s) \approx \int_0^1 \dots \int_0^1 \!  \left(  \prod_{i,j}   \mathrm{d}\nu_{ij} \right) e^{n\left[\sum_{ij} \nu_{ij} (\log(\sum_{k} \nu_{ik}) - \log(\nu_{ij})) +  \sum_{ij} \left(s_{ij} + \log \Pi_{ij} \right)\  \nu_{ij}\right]} \left(\prod_{i} \delta_{\sum_{j} \nu_{ji},\sum_{j} \nu_{ij}}\right) \delta_{\sum_{ij} \nu_{ij}, 1}\  .
\end{equation}
We note that the integrand in \eqref{eq:largenmgf} can be brought to the form $e^{n \lambda_N[\nu]}$, with the following definitions:

\begin{eqnarray}
    \label{eq:FinAct}
    &\lambda_N  &[\nu] = \lambda_1[\nu]+ \lambda_2[\nu]+ \lambda_3[\nu]+ \lambda_4[\nu]\\
    \label{eq:l1}
    &\lambda_1&[\nu] = \sum_{i=1}^N\sum_{j=1}^N \nu_{ij} \left( \log\left(\sum_{k=1}^N \nu_{ik}\right) - \log(\nu_{ij}) \right) \\
    \label{eq:l2}
    &\lambda_2&[\nu] = \sum_{i=1}^N \sum_{j=1}^N \log (\Pi_{ij}) \  \nu_{ij}\\
    \label{eq:l3}
    &\lambda_3&[\nu] = \sum_{i=1}^N \sum_{j=1}^N s_{ij} \ \nu_{ij}\\
    \label{eq:l4}
    &\lambda_4&[\nu] = \epsilon \left( \sum_{i=1}^N \sum_{j=1}^N \nu_{ij} - 1 \right)   + \sum_{i=1}^N \eta_i \left(\sum_{j=1}^N \nu_{ij}- \sum_{j=1}^N \nu_{ji}\right) \ ,
\end{eqnarray}
where $\epsilon$ and $\eta_i$ are Lagrange multipliers fixing the respective constraints. In \eqref{eq:FinAct} each term has a clear physical interpretation: $\lambda_1$, in \eqref{eq:l1}, is the geometric -- viz.\ related to the connectivity of the graph $\mathbf{G}$ -- entropy of a random walk on a graph with nodes and links contributions, akin to the entropy of a free particle; $\lambda_2$, in \eqref{eq:l2}, is the entropy due to the dynamics, encoded in the transition matrix; $\lambda_3$, in \eqref{eq:l3}, is the tilting potential necessary to drive the system towards a fluctuation of the pair empirical occupation measure; finally, $\lambda_4$ in \eqref{eq:l4}, enforces the normalisation and Kirchhoff-law (global balance). 

We can calculate the leading order in $n$ of \eqref{eq:largenmgf} via a saddle-point approximation, arriving at
\begin{equation}
    Z_{N,n}(s) \approx e^{n\lambda_N[\nu^*]} \ , 
\end{equation}
where $\nu^* = \text{argmin}_{\nu, \epsilon, \eta} \ \lambda_N[\nu]$ and $\nu^*$ are the minimisers of $\lambda_N[\nu]$ with respect to the set of $\nu_{ij}$s, $\eta_i$s and $\epsilon$. From the Euler--Lagrange equations for critical points of \eqref{eq:FinAct}, we find the following implicit expression for $\nu_{ij}^*$:
\begin{equation}
    \label{eq:saddle}
    \nu_{ij}^* =  \left( \Pi_s \right)_{ij} \left(\frac{e^{-\eta_j}}{e^{-\epsilon} e^{-\eta_i}} \right) \sum_{k=1}^N \nu_{jk}^* \ ,
\end{equation}
where the tilted matrix introduced in \eqref{eq:TiltMatr} appears. From \eqref{eq:saddle} we can write self-consistent conditions for $\epsilon$ and $\eta_i$ as follows:

\begin{eqnarray}\label{eq:righteigenvectors}
\sum\limits_j \left( \Pi_s \right)_{ij} e^{-\eta_j} &=& e^{-\epsilon} e^{-\eta_i} \\
\label{eq:lefteigenvectors}
\sum\limits_i \left( \Pi_s\right)_{ij} \frac{\sum_k \nu_{ik}^*}{e^{-\eta_i}} &=& e^{-\epsilon} \frac{\sum_k \nu_{jk}^*}{e^{-\eta_j} } \ ,
\end{eqnarray}

which reveal that $e^{-\epsilon}$ is an eigenvalue of the tilted matrix $\Pi_s$ with right eigenvector components $r_i = e^{-\eta_i}$ and left eigenvector components $l_j = \sum_k \nu_{jk}^*/e^{-\eta_j} $. Substituting \eqref{eq:righteigenvectors} into \eqref{eq:FinAct} we get
\begin{equation}\label{eq:connectionspectral}
    \lambda_N[\nu^*] = -\epsilon \ ,
\end{equation}
and, in particular, since $\lambda_N[\nu^*]$ is a maximum, $e^{\epsilon}$ is the dominant eigenvalue of \eqref{eq:TiltMatr}. The same conclusion can be reached by noticing that the left and right eigenvector elements in \eqref{eq:righteigenvectors} and \eqref{eq:lefteigenvectors} are all positive, which is true only for the dominant eigenvalue. These arguments provide a direct link with spectral methods. In particular, \eqref{eq:connectionspectral} provides an expression for the logarithm of the dominant eigenvalue of the tilted matrix. 

\newtext{Remarkably, this approach also provides an alternative expression for the so-called driven (or effective) process. This is a modified Markov chain that explains how specific fluctuations are created in time \cite{Jack2010,Chetrite2013,Chetrite2015,Chetrite2015a}; under certain conditions, it is equivalent to the original Markov chain conditioned to visiting the fluctuation of interest. Useful spectral and variational expressions of the driven process already appeared in the papers just mentioned. Here, we offer another explicit variational representation valid for discrete-time Markov chains. In agreement with \cite{Chetrite2015a}, the minimisers $\nu^* = \left\lbrace \nu^*_{ij} \right\rbrace$ of the action functional \eqref{eq:FinAct} characterise the driven process transition matrix with components
\begin{equation}
    \label{eq:DrivenProcess}
    \tilde{\Pi}_{ij} = \frac{\nu^*_{ij}}{\sum_{k=1}^N \nu^*_{ik}} \ .
\end{equation}
This last expression offers an alternative way to physically study and simulate the appearance of fluctuations and rare events in discrete-time Markov chain models.}

Concluding, in \eqref{eq:FinAct} we have obtained $\lambda_N$, the SCGF associated with the probability distribution of the pair empirical occupation measure in \eqref{eq:PairEmp}. To get the rate functional \eqref{eq:Rate} we only need to Legendre--Fenchel transform the SCGF in \eqref{eq:FinAct}, i.e.,
\begin{equation}
    \label{eq:LegFenSCGF}
    \sup_{s} \left( \sum_{i=1}^N \sum_{j=1}^N s_{ij} \nu^*_{ij} - \lambda_N[\nu^*] \right) = \sup_{s} \left( \lambda_3[\nu^*] - \lambda_N[\nu^*] \right) =  - \lambda_1[\nu^*] - \lambda_2[\nu^*] -\lambda_4[\nu^*] = H[\nu^*] \ ,
\end{equation}
where in the last step we recognise the pair empirical rate functional (with the necessary constraints -- mentioned and understood in \eqref{eq:Rate} -- fixed by the Lagrange multipliers in $\lambda_4$).

Assuming that one is interested in studying large fluctuations of an observable of the form \eqref{eq:Cost}, we remark that the associated SCGF can be obtained simply replacing $\lambda_3[\nu]$ in \eqref{eq:l3} with
\begin{equation}
    \lambda_3[\nu]  = s \sum_{i,j=1}^N f(i,j) \nu_{ij} \ , 
\end{equation}
where $s$ is the tilting parameter conjugated to $C_n$. \newtext{For instance, in physics applications, it is often of interest to consider the empirical current $\mathbb{J}_n(i,j) = L_n^{(2)}(i,j) - L_n^{(2)}(j,i)$, viz.\ the antisymmetric part of the pair-empirical occupation measure in \eqref{eq:PairEmp}, or again the occupation measure itself $L_n(i) = \sum_{j=1}^N L_n^{(2)}(i,j)$. The empirical current is an important observable as it allows us to estimate how far a system lies from equilibrium, whereas the occupation measure gives an estimate of the time spent by the system in each state of the state space.}

\newtext{
\section{Two-state model} \label{sec:example}

In this Section, in order give a more pedagogical understanding of how one could use \eqref{eq:permutationperformed} to derive leading, i.e., the SCGF in \eqref{eq:FinAct}, and finite $n$ behaviour, we compare our method with the more standard spectral approach on a simple two-state Markov chain. We show that the two methods give equivalent results and propose a physical interpretation of all terms appearing in the SCGF. We consider a general two-state Markov chain, whose transition matrix $\Pi$ reads

\begin{equation}
\Pi = 
    \begin{pmatrix}
    1 - p & p \\
    q & 1 - q
    \end{pmatrix} \ ,
\end{equation}
with $p$ and $q$ between $0$ and $1$. We choose to observe the flux between node  $1$ and node $2$, that is

\begin{equation}\label{eq:Observable2states}
    C_n = \frac{1}{n}\sum_{\ell = 1}^{n} \delta_{X_{\ell},1}\delta_{X_{\ell+1},2}  = \frac{t_{12}}{n}\ .
\end{equation}

The long-time behaviour of $C_n$ is given by $\lim_{n \rightarrow \infty} C_n = \frac{pq}{p+q} \eqqcolon c^*$. Intuitively, when $t_{12}$ is large, the Markov chain jumps frequently from $1$ to $2$ and from $2$ to $1$; instead, when $t_{12}$ is small the chain spends most of the time jumping from $1$ to $1$ and/or from $2$ to $2$. This situation is reminiscent of a particle in a double well potential immersed in a thermal bath, where temperature---that is, the strength of noise---regulates the frequency of jumps between the two minima. In this two-state model, the tilting parameter $s$ plays a role analogous to the temperature.

\subsection{Spectral approach}

The moment generating function can be computed using spectral methods. We start from \eqref{eq:moment_trajectories} (restricted to the case of the observable \eqref{eq:Observable2states}), i.e.,
\begin{equation}
    Z_{N,n}(s) =  \sum_{X_1, \dots, X_{n+1}}   \mathbb{P}(X_1) \prod_{\ell=1}^{n} \Pi_{X_{\ell}, X_{\ell+1}} e^{ s \delta_{X_{\ell},1} \delta_{X_{\ell +1},2}} \ ,
\end{equation}
which can be cast in the form
\begin{equation}
    Z_{N,n}(s) = \langle \mathbb{P}_1 | \left(\Pi_s\right)^n | 1 \rangle  \ ,
\end{equation}
where $\langle \mathbb{P}_1 |= (1,0)$ is the vector of initial probabilities, $| 1 \rangle = (1,1)$ and $\Pi_s$ is the tilted matrix, viz.\ \eqref{eq:TiltMatr} restricted to the case at hand, which reads

\begin{equation}
\Pi_s = 
    \begin{pmatrix}
    1 - p & p e^s \\
    q & 1 - q
    \end{pmatrix} \ .
\end{equation}
We can use the spectral decomposition of $\Pi_s$ to get

\begin{equation}
    Z_{2,n}(s) = \langle P_0 |\left( | r^+ \rangle \langle l^+| \Lambda_+^n +  | r^- \rangle \langle l^-| \Lambda_-^n \right)| 1 \rangle  = r_1^+ \Lambda_+^n (l_1^+ + l_2^+) + r_1^- \Lambda_-^n (l_1^- + l_2^-) \ ,
\end{equation}
where $\Lambda_{\pm}$ are the eigenvalues of $\Pi_s$ and $l^{\pm}$, $r^{\pm}$ the corresponding left and right eigenvectors, respectively. We notice that---for the spectral decomposition of $\Pi_s$ to be valid---left and right eigenvectors have to be bi-orthonormal. 

\begin{figure}
    \centering
    \includegraphics[scale=0.45]{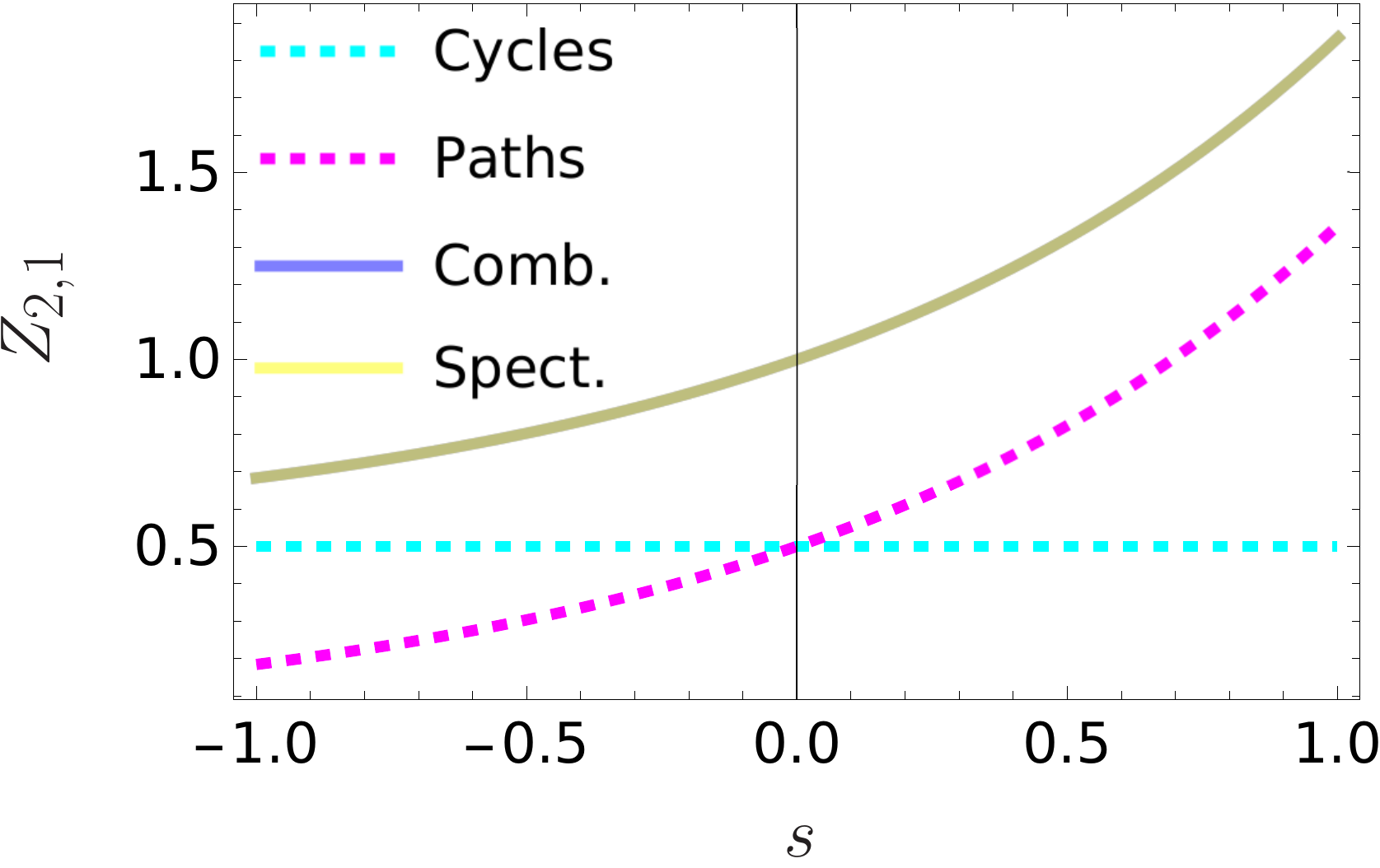} 
    \includegraphics[scale=0.45]{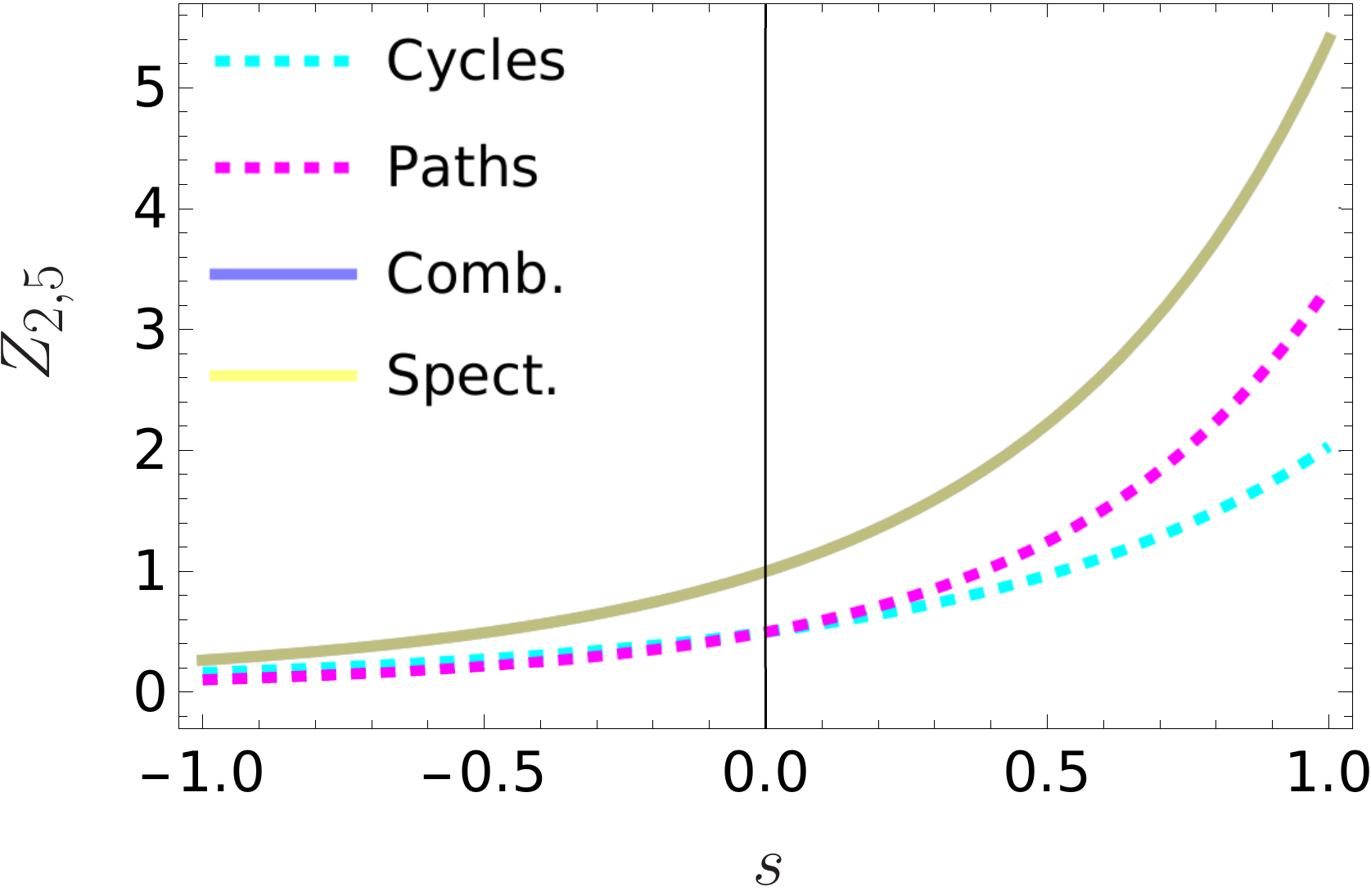} \\
    \includegraphics[scale=0.45]{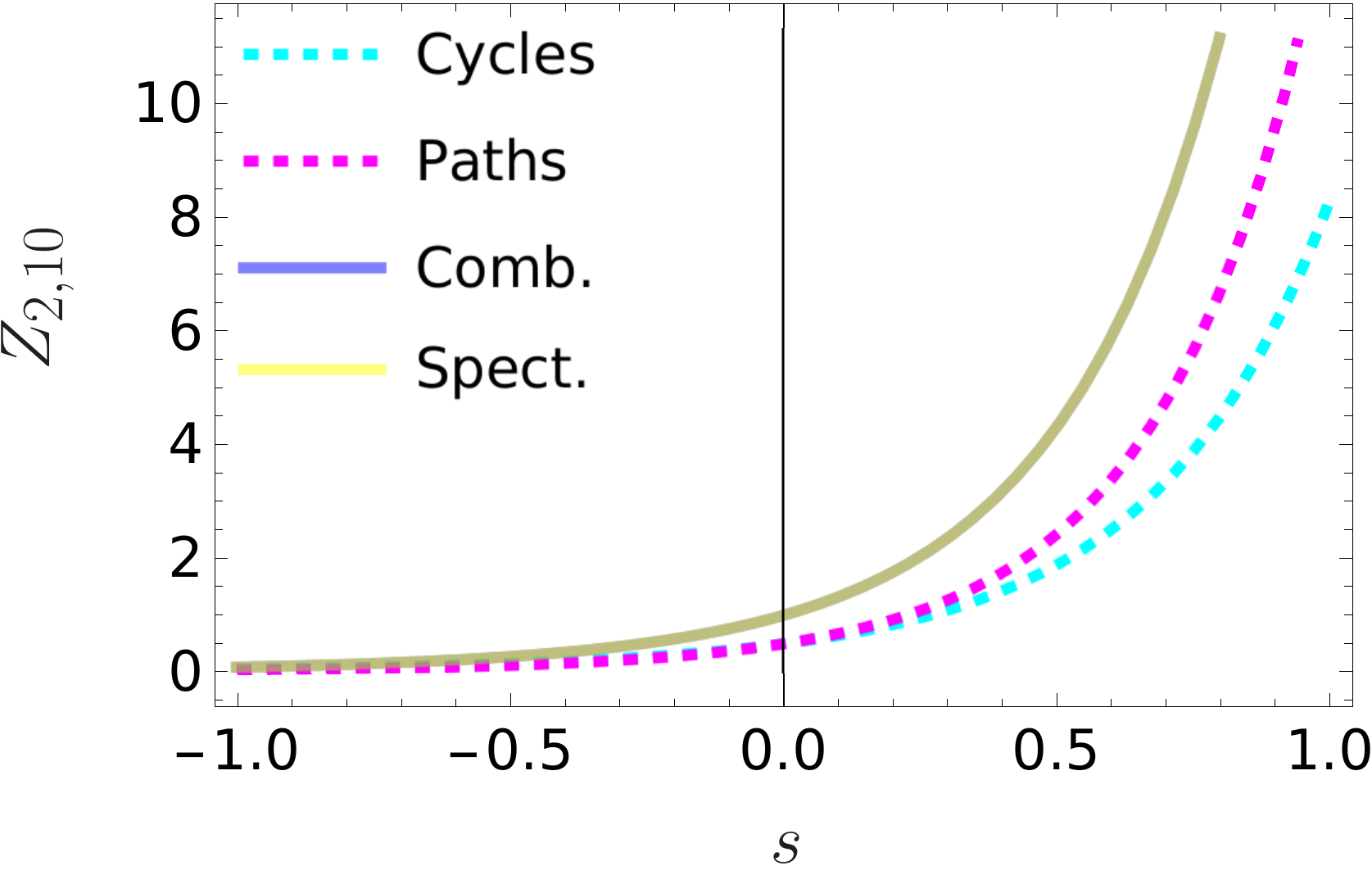} 
    \includegraphics[scale=0.45]{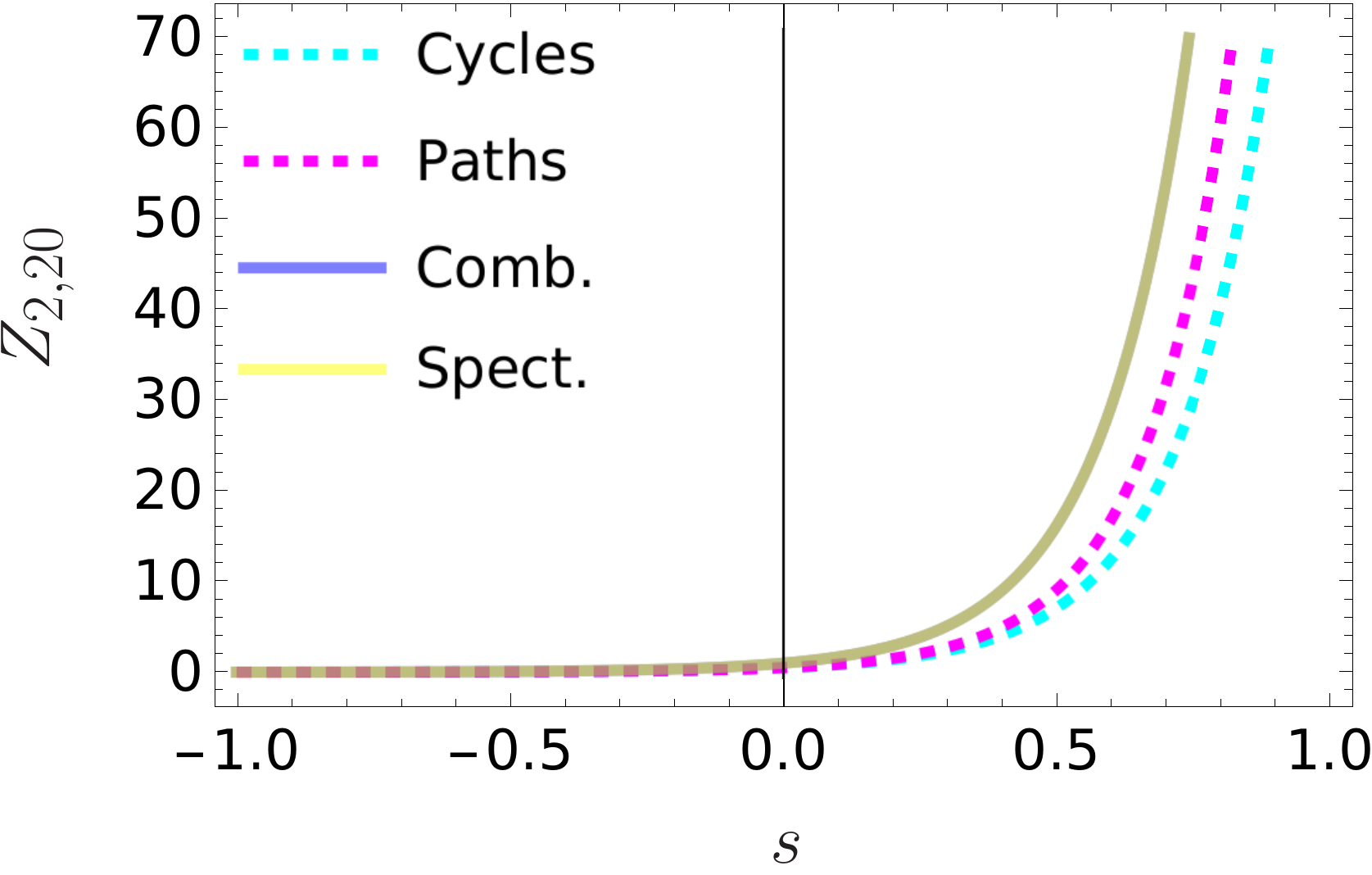}
    \caption{From top left to bottom right: Cycles and Paths contributions to the moment generating function, their sum (Comb.) obtained using the graph-combinatorial approach, in comparison with the moment generating function obtained via the spectral decomposition (Spect.) for increasing values of $n$. Comb.\ and Spect.\ curves fully overlap and the green-ish color shown is obtained by the combination of blue (Comb.) and yellow (Spect.). Furthermore, as expected, the Cycles contribution to the moment generating function is smaller for $s>0$ and larger for $s<0$ than the Paths one. To generate these plots, we have used $p = q = 1/2$.}
    \label{fig:2StateMoment}
\end{figure}

By computing the eigenvalues and eigenvectors of $\Pi_s$ explicitly, we arrive at
\begin{align}
\label{eq:2stateMomentSpectral}
\begin{split}
    & Z_{2,n}(s) =  \frac{1}{2^{n+1} \sqrt{(p-q)^2 + 4 pq e^s}} \times \\
    &\hspace{1cm} \times \Bigg\{ \left( (1 - 2 e^s) p - q \right) \left[ \left(2 - p - q - \sqrt{(p-q)^2 + 4 pq e^s}\right)^n - \left(2 - p - q + \sqrt{(p-q)^2 + 4 pq e^s}\right)^n \right]  \\
    &\hspace{1cm}  \left[ \sqrt{(p-q)^2 + 4 pq e^s} \left(2 - p - q - \sqrt{(p-q)^2 + 4 pq e^s}\right)^n + \left(2 - p - q + \sqrt{(p-q)^2 + 4 pq e^s}\right)^n \right] \Bigg\}  \ .
\end{split}
\end{align}

\subsection{Graph-combinatorial approach}

The moment generating function can also be computed using \eqref{eq:permutationperformed}. Remarkably, this other formulation highlights two different contributions coming from cycles and paths traveled starting from state $1$ of the state space. These can explicitly be written as
\begin{align}
    \label{eq:2StateCycles}
    Z_{2,n}^{\text{Cycles}}(s) &= (1-p)^n + \sum_{t_{11}=0}^n \sum_{t_{12}=1}^n \sum_{t_{22}=0}^n \delta_{t_{11} + 2 t_{12} + t_{22},n} \left( p q e^s \right)^{t_{12}} (1-p)^{t_{11}} (1-q)^{t_{22}} {t_{11} + t_{12} \choose t_{11}} {t_{22} + t_{12} - 1 \choose t_{22}} \\
    \label{eq:2StatePaths}
    Z_{2,n}^{\text{Paths}}(s) &= \sum_{t_{11}=0}^n \sum_{t_{12}=1}^n \sum_{t_{22}=0}^n \delta_{t_{11} + 2 t_{12} + t_{22} - 1,n} \frac{\left( p q e^s \right)^{t_{12}}}{q} (1-p)^{t_{11}} (1-q)^{t_{22}} {t_{11} + t_{12} - 1 \choose t_{11}} {t_{22} + t_{12} - 1 \choose t_{22}} \ ,
\end{align}
where in $Z_{2,n}^{\text{Cycles}}(s)$ we made explicit the cycle contribution coming from staying for $n$ consecutive steps on state $1$, and in $Z_{2,n}^{\text{Paths}}(s)$ the counting needs to start from $t_{12}=1$ because to have a meaningful path contribution the Markov chain needs to hop at least once from state $1$ to state $2$. We remark that in this simple model $\det (L_1) = t_{12}$ if $t_{12} \neq 0$ (in such a case the term is absorbed in \eqref{eq:2StateCycles} and \eqref{eq:2StatePaths} by the binomial coefficients), while when $t_{12} = 0$ the Laplacian is a $1 \times 1$ matrix: $L_1$ is thus an empty matrix, and we take its determinant to be $1$ for consistency.

We can find an explicit expression for \eqref{eq:2StateCycles} and \eqref{eq:2StatePaths} analytically. We replace the delta functions appearing by their contour integral representations

\begin{equation}
    \delta_{i,j} = \frac{1}{2\pi i} \oint_{|z|=1} z^{i-j-1} \mathrm{d} z \ .
\end{equation}

After making the substitution, we notice that the integrands in \eqref{eq:2StateCycles} and \eqref{eq:2StatePaths} are analytic functions everywhere except in $0$. This allows us to deform the integration contour to a circle of radius $\epsilon \ll 1$. The reason for this is to avoid spurious poles in the following steps.

We now let all the sums run up to $\infty$. This procedure is allowed as higher order terms in the sums do not affect the residue in $0$. The infinite sums can be explicitly evaluated and by doing so we get
\begin{align}
    \label{eq:2StateCyclesC}
    Z_{2,n}^{\text{Cycles}}(s) &= (1-p)^n + \frac{1}{(e^s p q - pq + q+ p-1)} \frac{1}{2 \pi i} \oint_{|z|=\epsilon} \frac{1}{z^{n-1}} \frac{e^s p q}{(z(1-p)-1)(z-z^*_1)(z-z^*_2)} \\
    \label{eq:2StatePathsC}
    Z_{2,n}^{\text{Paths}}(s) &= - \frac{1}{(e^s p q - pq + q+ p-1)} \frac{1}{2 \pi i} \oint_{|z|=\epsilon} \frac{1}{z^n} \frac{e^s p}{(z-z^*_1)(z-z^*_2)} \ ,
\end{align}
where
\begin{align}
    \label{eq:z1*}
    z^*_1 &= \frac{2}{2-p-q-\sqrt{(p-q)^2+4 e^s p q}} \\
    \label{eq:z2*}
    z^*_2 &= \frac{2}{2-p-q+\sqrt{(p-q)^2+4 e^s p q}} \ .
\end{align}
Notice that $z^*_1$ and $z^*_2$ are exactly the inverse of the eigenvalues found with spectral methods. We remark that the integrands in \eqref{eq:2StateCyclesC} and \eqref{eq:2StatePathsC} have acquired new singularities, in the form of simple poles at $z^*_1$, $z^*_2$ and $1/(1-p)$: these poles are unphysical, in the sense that their residue should not be considered when computing the contour integrals.

\begin{figure}
    \centering
    \includegraphics[scale=0.45]{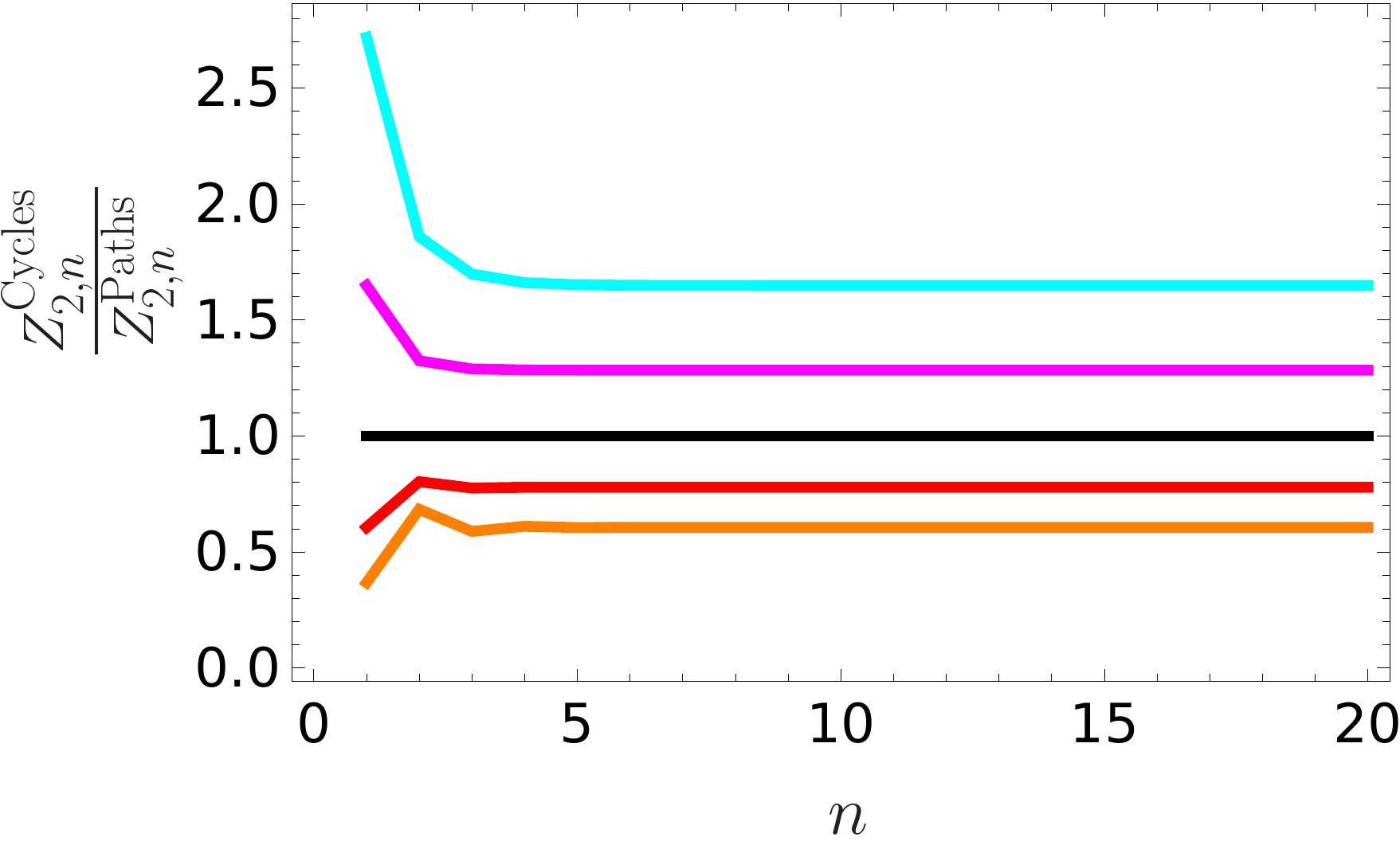}
    \caption{Ratios $\frac{Z_{2,n}^{\text{Cycles}}}{Z_{2,n}^{\text{Paths}}}$ as a function of $n$ for five different values of $s$, which are, from top to bottom: cyan, $s = -1.0$; magenta, $s = -0.5$; black, $s=0$; red, $s= 0.5$; orange, $s=1.0$}  
    \label{fig:2StateRatio}
\end{figure}

We can express $Z_{2,n}^{\text{Cycles}}(s)$ and $Z_{2,n}^{\text{Paths}}(s)$ as
\begin{align}
    \label{eq:2StateCyclesR}
    Z_{2,n}^{\text{Cycles}}(s) &= (1-p)^n + \frac{1}{(e^s p q - pq + q+ p-1)} \text{Res}_{z=0} \left( \frac{1}{z^{n-1}} \frac{e^s p q}{(z(1-p)-1)(z-z^*_1)(z-z^*_2)} \right) \\
    \label{eq:2StatePathsR}
    Z_{2,n}^{\text{Paths}}(s) &= - \frac{1}{(e^s p q - pq + q+ p-1)} \text{Res}_{z=0} \left( \frac{1}{z^n} \frac{e^s p}{(z-z^*_1)(z-z^*_2)} \right) \ .
\end{align}
Computing the residues we find 
\begin{align}
\label{eq:2StateCyclesFin}
Z_{2,n}^{\text{Cycles}}(s) &= (1-p)^n + \\
&\hspace{-0.5cm}+ \frac{(e^s p q (z_2^*)^{-n} (z_1^* ((z_2^*/z_1^*)^n - (z_2^* - p z_2^*)^n) + z_2^* (-1 + (z_2^* - p z_2^*)^n + (-1 + p) z_1^* (-1 + (z_2^*/z_1^*)^n))))}{(e^s p q - p q + p + q - 1)((1 + (-1 + p) z_1^*) (z_1^* - z_2^*) (1 + (-1 + p) z_2^*))} \\
\label{eq:2StatePathsFin}
Z_{2,n}^{\text{Paths}}(s) &= \frac{(e^s p (z_2^*)^{-n} (-1 + (z_2^*/z_1^*)^n))}{(e^s p q - pq + q+ p-1)(z_1^* - z_2^*)} \ .
\end{align}
By summing these two contributions and replacing $z^*_1$ and $z^*_2$ from \eqref{eq:z1*} and \eqref{eq:z2*}, we obtain exactly \eqref{eq:2stateMomentSpectral}.

In Fig.\ \ref{fig:2StateMoment} we show the functions $Z_{2,n}^{\text{Cycles}}$ and $Z_{2,n}^{\text{Paths}}$ and compare them with the moment generating function previously obtained via spectral methods.

Evidently, the moment generating function obtained by summing up cycles and paths contributions completely matches the moment generating function obtained with spectral methods, as the two curves are indistinguishable. An advantage of the graph-combinatorial approach with respect to the spectral calculation is the possibility to split the contributions coming from cycles and paths. As expected for the simple model investigated, cycles contribute less to the moment generating function for $s>0$ with respect to paths, and viceversa for $s<0$. The reason for this is that in the path scenario the Markov chain has to jump at least once from $1$ to $2$, contributing to $C_n$. The larger the $n$, the less pronounced is this effect. We also show in Fig.\ \ref{fig:2StateRatio} the ratio $Z_{2,n}^{\text{Cycles}}/Z_{2,n}^{\text{Paths}}$ for a few fixed values of the tilting parameter $s$ as a function of time $n$. Noticeably, the ratios become constant for $n$ big enough, supporting the fact that both $Z_{2,n}^{\text{Cycles}}$ and $Z_{2,n}^{\text{Paths}}$ share the same asymptotics for large $n$ and differ only by a constant prefactor that is a function of $s$.

\subsection{Large deviation regime}

We now investigate fluctuations in the large-$n$ limit computing the SCGF $\lambda(s)$. We compare the spectral and the variational formulae, to highlight the benefits of both approaches.

Using the spectral approach, the logarithm of the dominant eigenvalue is the SCGF \eqref{eq:2stateMomentSpectral} and reads
\begin{equation}
    \label{eq:2stateSCGF}
    \lambda(s) = \log \frac{2 - p - q + \sqrt{(p-q)^2 + 4pq e^s}}{2} \ .
\end{equation}

We can arrive at the same result by minimising \eqref{eq:FinAct}. Noticing that Kirchhoff law reduces to $\nu_{12} = \nu_{21}$, the action functional reduces to

\begin{figure}
     \centering
     \begin{subfigure}[b]{0.48\textwidth}
         \centering
         \includegraphics[width=\textwidth]{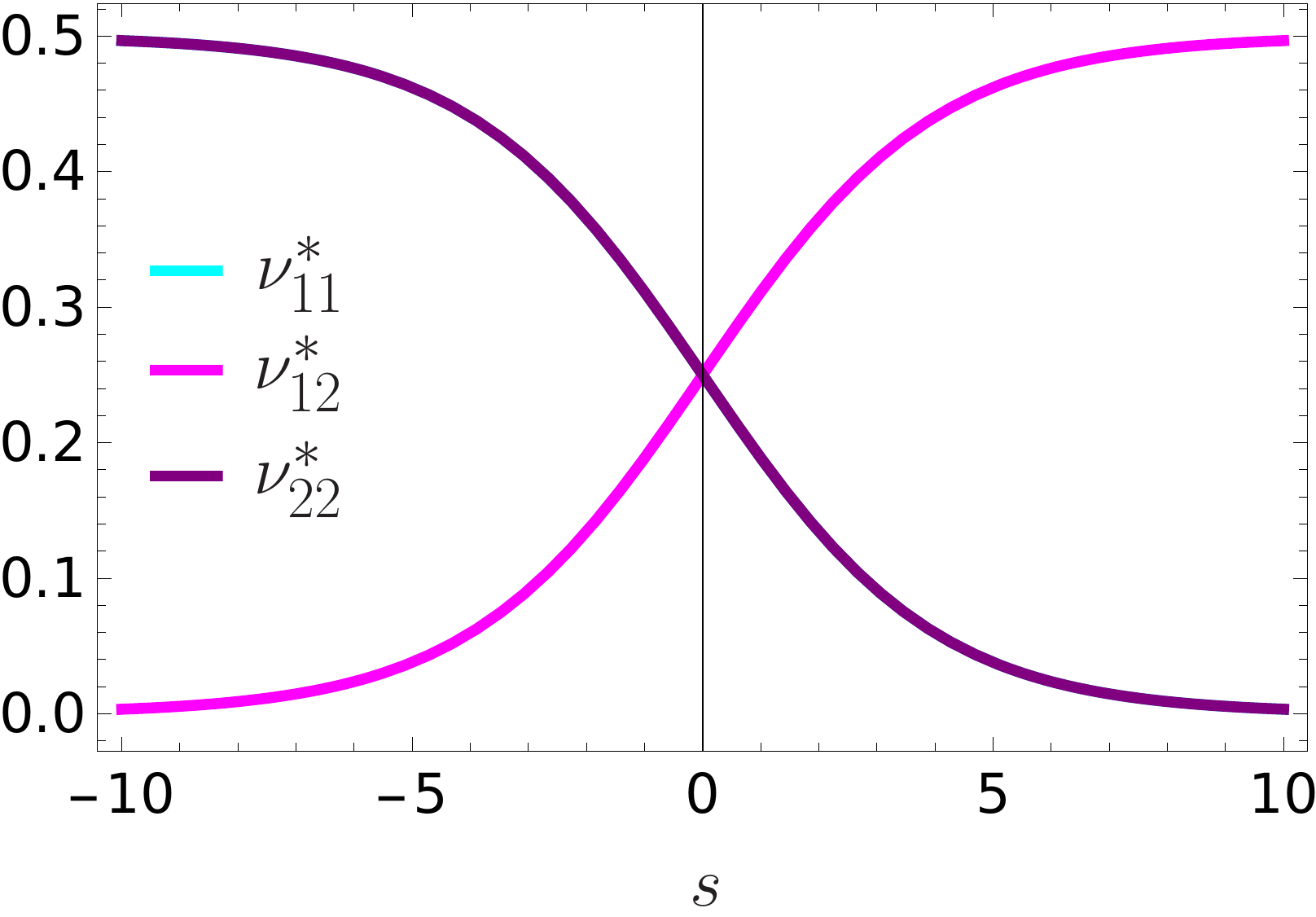}
         
         \label{fig:minimizers}
     \end{subfigure}
     \hfill
     \begin{subfigure}[b]{0.48\textwidth}
         \centering
         \includegraphics[width=\textwidth]{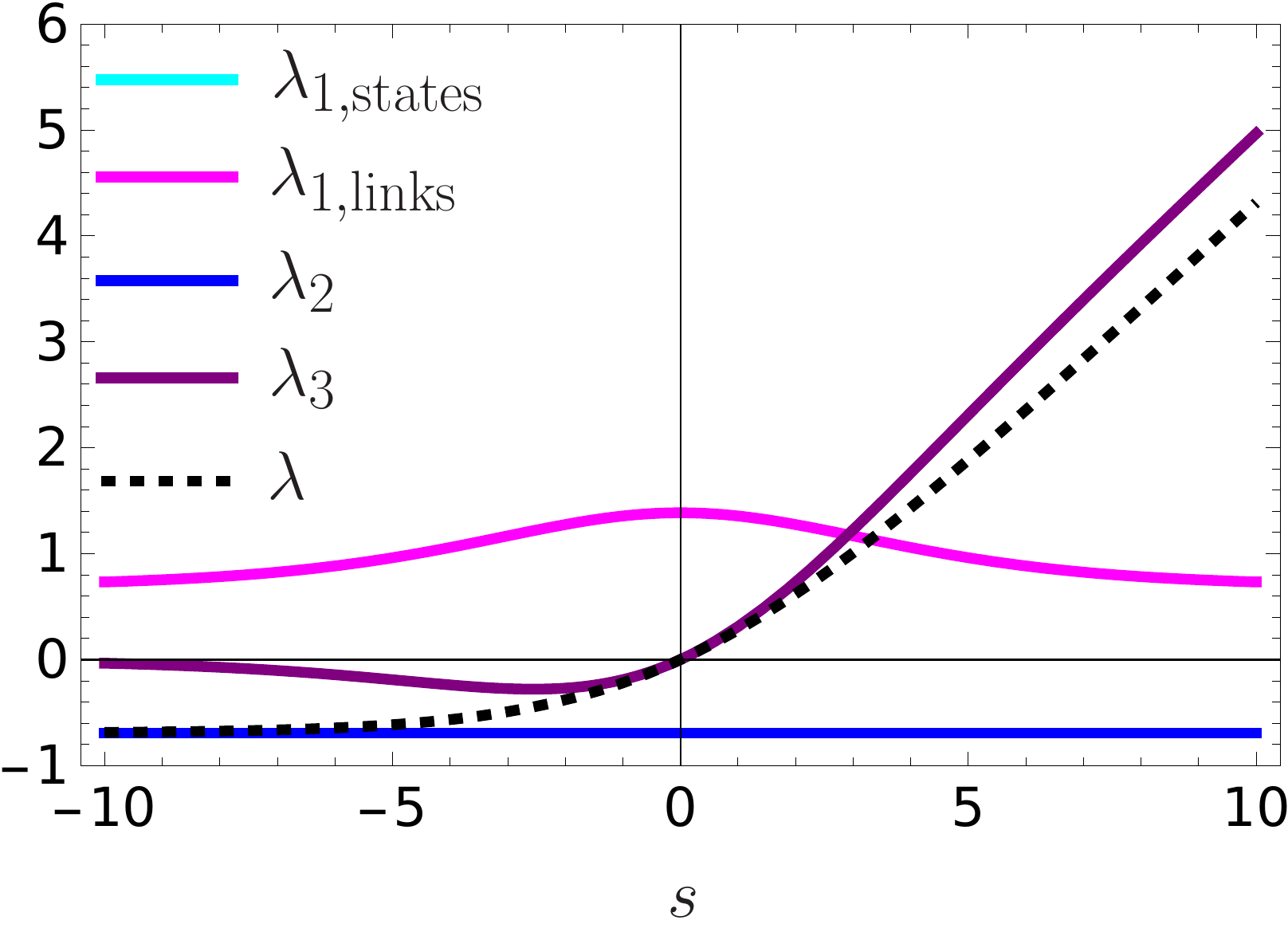}
        
         \label{fig:actionpieces}
     \end{subfigure} \\
     
     \begin{subfigure}[b]{0.48\textwidth}
         \centering
         \includegraphics[width=\textwidth]{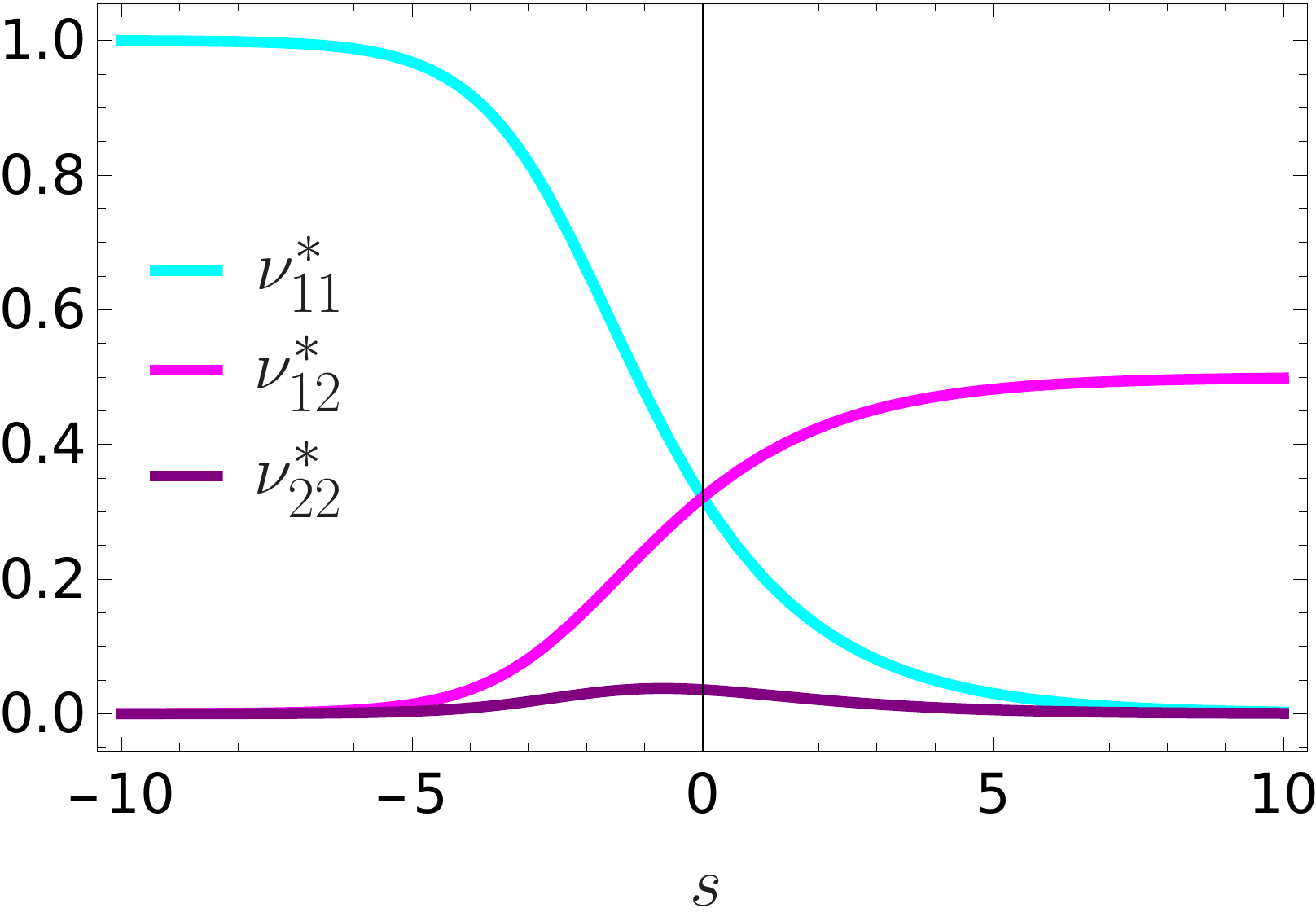}
         
         \label{fig:minimizersLoc}
     \end{subfigure}
     \hfill
     \begin{subfigure}[b]{0.48\textwidth}
         \centering
         \includegraphics[width=\textwidth]{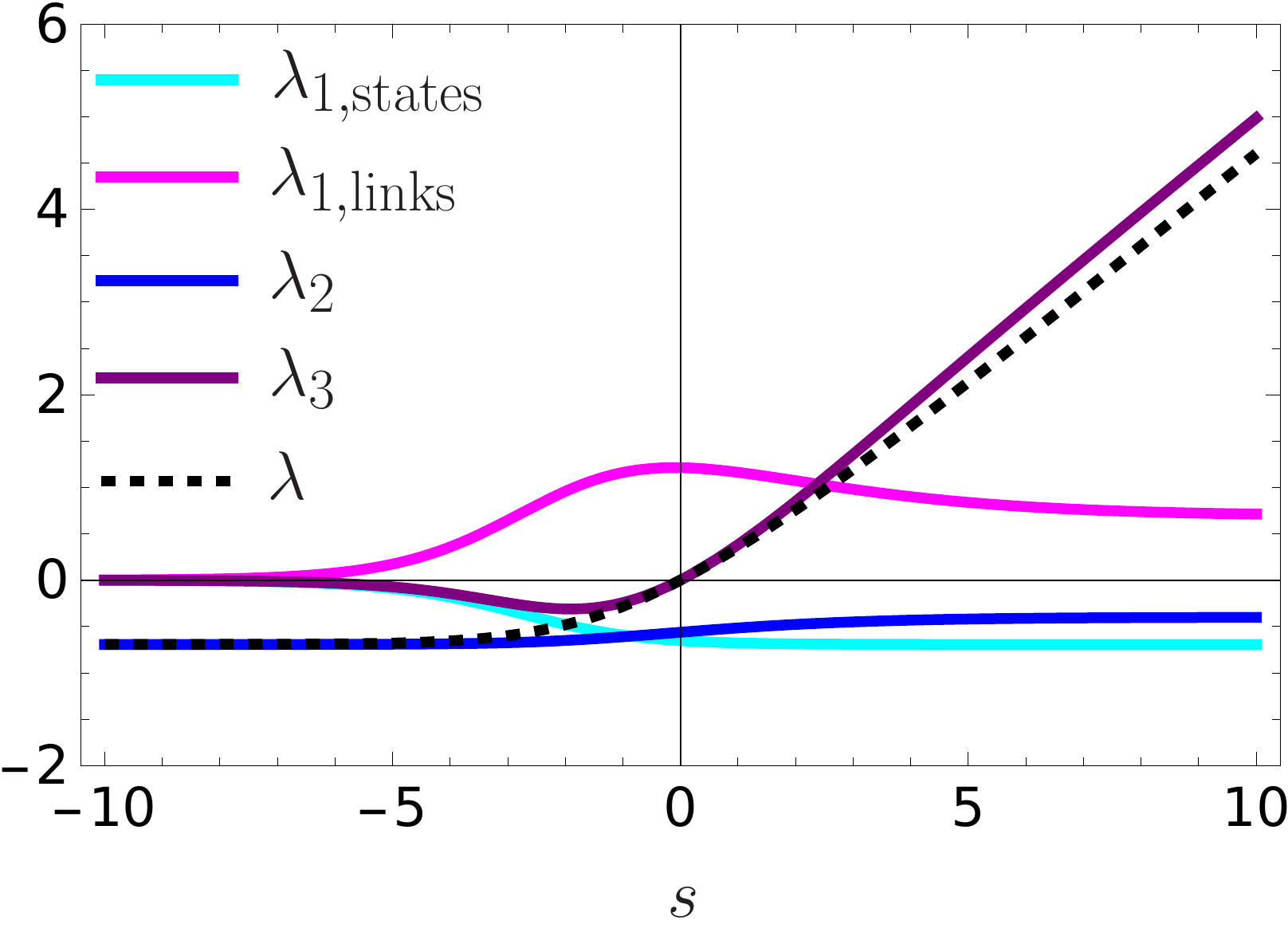}
        
         \label{fig:actionpiecesLoc}
     \end{subfigure}
     
    \caption{Top-left panel: Plot of the minimizers $\nu^*_{11}$, $\nu^*_{12}$, $\nu^*_{22}$ of the form \eqref{eq:minimisernu11}, \eqref{eq:minimisernu12}, \eqref{eq:minimisernu22} for $p = q = 0.5$ as a function of $s$. We notice that the curve of $\nu^*_{11}$ coincides with that of $\nu^*_{22}$, due to the symmetry of $p$ and $q$.  Top-right panel: Plot of all the contributions to the SCGF as defined in \eqref{eq:l2}, \eqref{eq:l3}, \eqref{eq:l4}, \eqref{eq:l1states}, \eqref{eq:l1links} of the two-state model and their sum as a function of $s$ for $p = q = 0.5$. For this choice of parameters, we notice that the curve of $\lambda_{1, \text{states}}$ coincides with that of $\lambda_2$, and interestingly they do not depend on $s$. Bottom-left panel: Plot of the minimizers $\nu^*_{11}$, $\nu^*_{12}$, $\nu^*_{22}$ of the form \eqref{eq:minimisernu11}, \eqref{eq:minimisernu12},  \eqref{eq:minimisernu22} for $p = 0.5$, $q = 0.9$ as a function of $s$. Bottom-right panel: Plot of all the contributions to the SCGF as defined in \eqref{eq:l2}, \eqref{eq:l3}, \eqref{eq:l4}, \eqref{eq:l1states}, \eqref{eq:l1links} of the two-state model and their sum as a function of $s$ for $p = 0.5$ and $q = 0.9$. }
    \label{fig:asymptotics2state}
    
\end{figure}

\begin{align}
\begin{split}\label{eq:action2states}
    &\lambda[\nu] = (\nu_{11} + \nu_{12}) \log(\nu_{11} + \nu_{12}) + (\nu_{12} + \nu_{22}) \log(\nu_{12} + \nu_{22}) - \log{\nu_{11}}  - \log{\nu_{22}} \\ &+
    \nu_{11}\log (1 -p) + \nu_{12}(\log p +  \log q) +
    \nu_{22}\log(1-q) + s \nu_{12} + \epsilon(\nu_{11} + 2\nu_{12} +  \nu_{22} -1 ) \ .
\end{split}
\end{align}
The determinant of the system of equations satisfied by the minimum of \eqref{eq:action2states}, which is linear in $\nu$, must be $0$ to have non-trivial solutions. This condition gives an equation for $e^{-\epsilon}$ ($\epsilon$ is the Lagrange multiplier fixing the normalisation condition) whose solution gives---through \eqref{eq:connectionspectral}---expression \eqref{eq:2stateSCGF}. This last can be replaced in the form of the minimisers obtained by solving the linear system, which read
\begin{align}
    \label{eq:minimisernu11}
    \nu^*_{11} &= \frac{(1-q)^{-1} e^{-\epsilon} - 1}{(1-q)^{-1} e^{-\epsilon} + 2 ((1-p)^{-1} e^{-\epsilon} - 1) ((1-q)^{-1} e^{-\epsilon} - 1) + (1-p)^{-1} e^{-\epsilon} - 2}\\
    \label{eq:minimisernu12}
    \nu^*_{12} &= \frac{((1-p)^{-1} e^{-\epsilon} - 1)((1-q)^{-1} e^{-\epsilon} - 1)}{(1-q)^{-1} e^{-\epsilon} + 2 ((1-p)^{-1} e^{-\epsilon} - 1) ((1-q)^{-1} e^{-\epsilon} - 1) + (1-p)^{-1} e^{-\epsilon} - 2}\\
    \label{eq:minimisernu22}
    \nu^*_{22} &= \frac{(1-p)^{-1} e^{-\epsilon} - 1}{(1-q)^{-1} e^{-\epsilon} + 2 ((1-p)^{-1} e^{-\epsilon} - 1) ((1-q)^{-1} e^{-\epsilon} - 1) + (1-p)^{-1} e^{-\epsilon} - 2} \ ,
\end{align}
to get their explicit form as a function of $p$, $q$, and the tilting parameter $s$.

In the top-left and bottom-left panels of Fig.\ \ref{fig:asymptotics2state} we plot the minimisers \eqref{eq:minimisernu11}, \eqref{eq:minimisernu12}, and \eqref{eq:minimisernu22} as a function of the tilting parameter $s$. For the top-left case, we use $p = q = 0.5$, while for the bottom case $p = 0.5$ and $q = 0.9$. We notice that when $p = q$, as in the top-left panel, $\nu^*_{11} = \nu^*_{22}$ identically. This reflects a permutation symmetry of the system: when $p=q$, switching states $1$ and $2$ does not affect the transition matrix. In this case, the Markov chain smoothly transitions between two regimes: for $s\ll0$, the chain spends half of the time in node $1$ and half on $2$; for $s\gg0$, the chain spends all the time jumping from state $1$ to state $2$ and back. When $p \neq q$, instead, for $s<0$ the system smoothly transitions to a localised state, where the Markov chain is mostly located on $1$ (resp. $2$) if $p<q$ (resp. $p>q$).
Interestingly, we notice that for $p < q$ the maximum of $\nu^*_{22}$ occurs at a finite and negative value of $s$.

In the top-right and bottom-right panels of Fig.\ \ref{fig:asymptotics2state} we plot each contribution to the SCGF obtained with our approach alongside their sum. The SCGF is in perfect agreement with the one obtained using spectral methods. Furthermore, our approach allows us to understand the magnitude of each physical term. We split $\lambda_1(s)$, as defined in \eqref{eq:l1}, into two terms as follows:
\begin{align}
    &\lambda_{1, \text{states}} = \sum_{i=1}^2\sum_{j=1}^2 \nu_{ij}  \log\left(\sum_{k=1}^2 \nu_{ik}\right)  \label{eq:l1states}\\
    & \lambda_{1,\text{links}} = -\sum_{i=1}^2 \sum_{j=1}^2 \nu_{ij} \log(\nu_{ij}) \label{eq:l1links}\ ,
\end{align}
and plot them separately. In the top-right panel we used $p = q = 0.5$, while in the bottom-right panel we used $p = 0.5$ and $q= 0.9$. In both cases, when $|s|$ is large we notice that $\lambda_{1, \text{states}}$ and $\lambda_{1,\text{links}}$ balance each other, and their sum is close to zero. This is because in both cases, the Markov chain spends most of the time in just a fraction of the available links. For $s \gg 0 $, the dominant contribution in both cases is due to the tilting term $\lambda_3(s)$. For $s\ll0$, we notice a striking difference: when $p = q$, both $\lambda_{1, \text{states}}$ and $\lambda_{1,\text{links}}$ tend to a finite value. This is because the chain still visits both node $1$ and node $2$. Instead, in the case $p \neq q$, $\lambda_{1, \text{states}}$ and $\lambda_{1,\text{links}}$ tend to $0$. This is because of the aforementioned localisation behaviour. In both cases, since the tilting term $\lambda_3(s)$ becomes negligible, the SCGF $\lambda$ is well approximated by $\lambda_2$, the dynamical entropy. 

}

\section{Conclusion}

In this work we propose a way to study the large deviation regime of fluctuations of two-point observables of a discrete-time Markov chain. Adopting graph-combinatorial arguments similar to those in \cite{Dawson1957,Goodman1958, Polettini2015}, we show how to calculate the finite-time moment generating function and the scaled cumulant generating function, objects that have a clear interpretation in the framework of statistical physics as they correspond, respectively, to the canonical partition function and Helmoltz free energy. In particular, all terms of the Helmoltz free energy have a clear physical meaning---see \eqref{eq:FinAct} and following discussion. \deletetext{A further benefit of our derivation is that it establishes} \newtext{We establish} a direct and explicit link with spectral methods, as the Lagrange multipliers in \eqref{eq:FinAct} can be shown to be the dominant eigenvalue and right eigenvector of the tilted matrix---see \eqref{eq:righteigenvectors}. \newtext{Furthermore, from the minimisers $\nu^*$ we show how to compute in a simple way the occupation measure on the nodes and the driven process.}

\newtext{We illustrate the benefits of our method in a general two-state model, for which we can compute analytically both the moment generating function and the SCGF. We show plots where we highlight the new information accessible with our method: in particular, we compare the different contributions of paths and cycles to the moment generating function. For the large deviation regime, analysing the minimisers $\nu^*$ as well as all the terms in our formula for the SCGF, we find an interesting localisation behaviour of the Markov chain when the two-state model is not symmetric.} 

Remarkably, the finite-time expression for the moment generating function could be used as the starting point for future investigations on the role of sub-leading terms in the fluctuations of observables, for which to our knowledge not much is known. \newtext{A remarkable contribution in this direction is \cite{Causer2022}, where authors use matrix products states to characterise fluctuations at finite time. An interesting avenue for future research would be to try to apply our methods in the continuous-time setting. } 

Furthermore, once we fix the state-space connectivity and probability weights, it would be interesting to understand the interplay between the long-time limit and the large number of states limit. In the framework of Markov chains satisfying detailed balance, this approach could, in principle, be adopted to investigate transient behaviour and metastability in rough energy landscapes, a problem relevant to many areas in statistical physics \cite{Baronchelli2009GlassTA,Arceri2020}. More generally, Markov chains that satisfy global balance but not detailed balance are a paradigmatic model for out of equilibrium phenomena. In this context, understanding finite-time behaviour is challenging---see \cite{Jia2014} for applications to biology. Out of equilibrium steady states are directly accessible in the large deviations framework \cite{Touchette2013,Cofre2019} and are of interest to many communities. 

Finally, we remark that the Helmoltz free energy associated with the pair empirical occupation measure is a powerful tool to investigate dynamical phase transitions in fluctuations of one and two-point observables. For instance, in \cite{DeBacco2016,Coghi2019} the authors show evidence of a localisation phase transition in random walks on random graphs. In an upcoming work, we intend to investigate toy models where this phenomenon can be analytically characterised using the approach outlined in this paper.

\section*{Acknowledgments}

GC and FC are thankful to Gianmichele Di Matteo for insightful discussions and to Mayank Shreshtha for having designed Fig.\ \ref{fig1}. FC is grateful to Hugo Touchette for pointing to interesting literature in the topic and for the hospitality in Stellenbosch (South Africa) during the writing stage of the manuscript. GC is supported by the EPSRC Centre for Doctoral Training in Cross-Disciplinary Approaches to Non-Equilibrium Systems (CANES, EP/L015854/1).

\bibliographystyle{ieeetr}
\bibliography{mybib}

\end{document}